\documentclass[12pt,reqno]{article}
\usepackage[lmargin=2cm, rmargin=2cm]{geometry}
\usepackage{amsfonts}
\usepackage{amssymb}
\usepackage{enumerate}
\usepackage{amsthm}
\usepackage{array}
\usepackage{epsfig}
\usepackage[small,bf]{caption}
\usepackage{appendix}
\usepackage{dsfont}
\usepackage{hyperref}
\usepackage{amsmath}
\usepackage{graphicx}
\usepackage{epstopdf}

\allowdisplaybreaks \numberwithin{equation}{section}

\begin{document}

\begin{titlepage}
 \thispagestyle{empty}

\begin{flushright}
    % \hfill{ITP-UU-14/29 }\\
     %  \hfill{CPHT-RR098.1214 }\\
 \end{flushright}

 \begin{center}

 \vspace{30mm}

     { \LARGE{\bf  {Entanglement in Fock space of \\ random QFT states}}}

     \vspace{40pt}

\Large{{\bf Javier M. Mag\'an and Stefan Vandoren}} \\[8mm]
{\small\slshape
Institute for Theoretical Physics \emph{and} Center for Extreme Matter and Emergent Phenomena, \\
Utrecht University, 3508 TD Utrecht, The Netherlands \\

\vspace{5mm}

{\upshape\ttfamily j.martinezmagan, s.j.g.vandoren@uu.nl}\\[3mm]}

\vspace{8mm}

     \vspace{10pt}

    \vspace{10pt}

\date{\today}

\end{center}

\begin{abstract}
Entanglement in random states has turned into a useful approach to quantum thermalization and black hole physics. In this article, we refine and extend the `random unitaries framework' to quantum field theories (QFT), and to include conserved charges. We show that in QFT, the connection between typical states, reduced subsystems and thermal dynamics is more transparent within the Fock basis. We provide generic formulae for the typical reduced density matrices and entanglement entropies of any given subset of particles. To illustrate our methods, we apply the generic framework to the simplest but non trivial cases, a massless scalar field in two dimensions and its generalization to the case of $N$ scalar fields, including the large $N$ limit. We find the effective temperature, by matching the reduced dynamics to a Gibbs ensemble, and derive the equation of state of the QFT. The deviations from perfect thermality are shown to be  of order $1/S$ instead of $\exp(-S)$, a result which might be relevant for black hole physics. Finally we describe the analogue of the so-called `Page curve' in the QFT scenario as a function of the energy scale which divides high from low energy degrees of freedom.
\end{abstract}

 \vspace{10pt}
\noindent

\end{titlepage}

\thispagestyle{plain}

\tableofcontents

\baselineskip 6 mm

\newpage

\section{Introduction and summary of results}

\subsection{Quantum thermalization and random matrices}

In its simplest formulation, the black hole information paradox \cite{paradox} follows from the statement that black hole radiation \cite{hawking} is precisely thermal, with the temperature depending just on the conserved charges defining the black hole. Given that there is an exponentially big number $\Omega=e^{S_{\textrm{BH}}}$ of initial states that could collapse to the given black hole, where $S_{\textrm{BH}}$ is the Bekenstein-Hawking entropy \cite{bekenstein,hawking}, it is concluded that after the black hole has evaporated the information about the initial state  is lost. Mathematically, the paradox arises since under unitary evolution of the initial quantum state, we can never produce a thermal ensemble,
\begin{equation}\label{iloss}
U\vert \Psi\rangle\langle\Psi\vert U^{\dagger} \neq \rho^{\textrm{Gibbs}}\;,
\end{equation}
where $\rho^{\textrm{Gibbs}}$ is an appropriate Gibbs distribution for the problem at hand. Interestingly, this problem is strikingly similar to one old problem in quantum mechanics, the problem of `quantum thermalization', see \cite{deutch, page, srednickisub} and references therein. The problem of quantum thermalization can be easily stated as:
\begin{itemize}
\item In the context of microscopic unitarity, and given~(\ref{iloss}), how can Gibbs ensembles emerge?
\end{itemize}
In what follows we will assume the Hilbert space to be factorizable\footnote{The ideas and results concerning the problem quantum thermalzation are most probably not sensitive to this condition. But in this framework the concepts can be explained more transparently, and a plethora of results can be derived. Besides, the models we will explicitly consider in this article fall in this factorizable class.} :
\begin{equation}\label{Hilbert}
\mathcal{H}=\mathcal{H}_{1}\otimes\mathcal{H}_{2}\otimes \cdots\otimes \mathcal{H}_{i}\otimes\cdots \;.
\end{equation}
The number of factors in~(\ref{Hilbert}) is unconstrained, it can be finite or infinite. Besides, the Hilbert space factors do not need to be equal to each other. If the algebra of operators acting on $\mathcal{H}_{i}$ is given by $\mathcal{O}_{i}^{a_{i}}$, with $a_{i}$ a discrete parameter, the algebra of operators acting on $\mathcal{H}$ is
\begin{equation}\label{operator}
\mathcal{O}=\mathcal{O}_{1}^{a_{1}}\otimes\mathcal{O}_{2}^{a_{2}}\otimes \cdots \mathcal{O}_{i}^{a_{i}}\otimes\cdots \;.
\end{equation}
Reduced subalgebras/subsystems $\mathbf{A}$ are defined by specifying a set of Hilbert factors, and considering non-unity operators only in those subfactors:
\begin{equation}\label{reducedoperator}
\mathcal{O}_{\textbf{A}}=\mathcal{O}_{\textrm{A}}\otimes \textbf{1}\otimes \textbf{1}\otimes \cdots\;.
\end{equation}
As is well known, given a unitarily evolving pure state $\rho(t)=\vert\psi (t)\rangle\langle\psi (t)\vert$, the expectation value of any operator belonging to such subalgebra $\textbf{A}$ is completely characterized by its associated reduced density matrix:
\begin{equation}\label{evolvedoperator}
\langle\psi (t)\vert\mathcal{O}_{\textbf{A}}\vert\psi (t)\rangle = \textrm{Tr}(\rho_{\textbf{A}}(t)\mathcal{O}_{\textrm{A}})\;,
\end{equation}
where
\begin{equation}\label{reduced}
\rho_{\textbf{A}}(t)=\textrm{Tr}_{\bar{\textbf{A}}}(\rho (t))\;,
\end{equation}
and $\bar{\textbf{A}}$ is the complement subalgebra/subsystem of $\textbf{A}$.

Although perfect thermality cannot be attained within unitary evolution, we might still expect certain approximate thermality for the correlation functions of the theory. The condition stating that all measurements~(\ref{evolvedoperator}) are approximately thermal
\begin{equation}\label{thermality1}
\langle\psi (t)\vert\mathcal{O}_{\textbf{A}}\vert\psi (t)\rangle = \textrm{Tr}(\rho^{\textrm{Gibbs}}\mathcal{O}_{\textbf{A}})\pm \textrm{error}\;,
\end{equation}
can be more rigorously stated as
\begin{equation}\label{thermality}
\rho_{\textbf{A}}(t)=\rho^{\textrm{Gibbs}}_{\textbf{A}}\pm \textrm{error}\;,
\end{equation}
with small errors, and where
\begin{equation}\label{reducedthermality}
\rho^{\textrm{Gibbs}}_{\textbf{A}}= \textrm{Tr}_{\bar{\textbf{A}}}(\rho^{\textrm{Gibbs}})\;,
\end{equation}
is the reduced thermal density matrix associated to $\textbf{A}$.

Given these preliminaries, some immediate questions concerning quantum thermalization are:
\begin{itemize}
\item What is the corresponding $\rho^{\textrm{Gibbs}}$ for the specific theory and state $\rho (t)$ considered? Or equivalently, how the parameters defining $\rho^{\textrm{Gibbs}}$ are related to the parameters defining $\rho (t)$?
\item How does the error scale in the thermodynamic limit?
\end{itemize}
Considering~(\ref{thermality}) and the previous questions, one modern approach to tackle this problem is to compute the entanglement entropy of the reduced subsystem $\textbf{A}$, which is defined by
 \begin{equation}\label{entanglement}
S_{\textrm{E}}^{\textbf{A}}(t)=-\textrm{Tr}(\rho_{\textbf{A}}(t)\log \rho_{\textbf{A}}(t))\;.
\end{equation}
If our state is such that relation~(\ref{thermality}) holds, the next insightful relation follows as well:
 \begin{equation}\label{entanglementthermal}
S_{\textrm{E}}^{\textbf{A}}(t)=S_{\textrm{Gibbs}}^{\textbf{A}}\pm \textrm{error}\;,
\end{equation}
which states that the origin of thermal entropy in quantum mechanics lies in quantum entanglement.

The discussion has been until now entirely generic. To proceed we should take a specific Hamiltonian $H$ and find the exact evolved quantum state:
\begin{equation}\label{evolved}
\rho(t)=\vert\psi (t)\rangle\langle\psi (t)\vert = U\vert\psi \rangle\langle\psi \vert U^{\dagger}=e^{-i H t}\vert\psi \rangle\langle\psi \vert e^{i H t}\;.
\end{equation}
This is certainly a difficult task, since we are usually interested in high energy states, with many particles excited\footnote{Some important progress has been done in the context of integrable systems, see \cite{integrable} and references therein, conformal field theories, see \cite{cft}, and in the context of holographic thermalization, see \cite{esperanza2010,vijay2011}. In this article we will be interested in other approach, which is more general in certain aspects and it is able to provide the scaling of the deviations from thermality with the size of the system.}. But due to these complicated dynamics, we expect that some approximations can be done. In particular, one common procedure is to assume that after a certain time we can approximate the evolved state as a \emph{random vector}\footnote{See  \cite{vijay} and \cite{harlow} for two nice expanded reviews on the subject.}:
\begin{equation}\label{randomvector}
\vert\psi (t\gtrsim t_{\textrm{relax}})\rangle \simeq   \vert\psi _{\textrm{random}}\rangle \;,
\end{equation}
which can be equivalently thought as being created by the application of a random unitary matrix to a given initial state: 
\begin{equation}\label{randomunitary}
\vert\psi _{\textrm{random}}\rangle = U_{\textrm{random}}\vert\psi \rangle  \;.
\end{equation}
The question now is obvious:
\begin{itemize}
\item How do we define $\vert\psi _{\textrm{random}}\rangle$ or $ U_{\textrm{random}}$?
\end{itemize}
If we could answer this question generically for any theory and initial state, it would allows us to compute typical values of physical quantities in a unitary framework. The physical quantities in which we are interested in this article are the typical\footnote{We use the words "typical"and "average" interchangeably. For example, the typical reduced density is by definition the same as the average density matrix.}  reduced density matrix of a given subsystem $\textbf{A}$, given by
\begin{equation}\label{typicalreduced}
\rho_{\textbf{A}}^{\textrm{typical}}=\int d\rho \,\textrm{Tr}_{\bar{\textbf{A}}}(\rho)\;,
\end{equation}
and the typical entanglement entropies
\begin{equation}\label{typicalentanglment}
S_{\textbf{A}}^{\textrm{typical}}=\int d\rho\, S_{\textrm{E}}^{\textbf{A}}(\rho)\;.
\end{equation}
This approach was pioneered in \cite{Lloyd, page}. In those articles a bipartite Hilbert space was considered, with dimension $AB$, where $A$ and $B$ are the dimensions of the corresponding factors of the total Hilbert space. For this Hilbert space, $\vert\psi _{\textrm{random}}\rangle$ was defined by averaging over the continuum manifold of quantum states \cite{Lloyd}, with a constant measure, i.e the Haar ensemble of quantum states in the full Hilbert space. Again, this is equivalent to applying a random unitary $ U_{\textrm{random}}$ picked from the Haar ensemble of random unitaries. With this specific definition of randomness in the given Hilbert space, the typical reduced density matrix~(\ref{typicalreduced}) was found in \cite{Lloyd} and, using those results, the typical entanglement entropy~(\ref{typicalentanglment}) was found in \cite{page}\footnote{See \cite{harlow},  and references therein, for an analogous but somewhat different approach coming from quantum information theory.}. In the light of the previous questions about quantum thermalization, their results for this specific case are the following:
\begin{itemize}
\item The typical reduced density matrix of a subsystem $\textbf{A}$ is proportional to the identity matrix. Therefore, this reduced density matrix might be associated to the thermal density matrix of any Hamiltonian (defined for the finite dimensional Hilbert space) in the infinite temperature limit.
\item The error depends on the size of both subsystems $\textbf{A}$ and $\textbf{B}$. For $A\simeq \mathcal{O}(1)$, the error in formula~(\ref{entanglementthermal}) is of $\mathcal{O}(\frac{1}{AB})\sim \mathcal{O}(2^{-S})$, where $S$ is the total entropy of the system. For $A=B$ the error is of $\mathcal{O}(1)$.
\end{itemize}
The first bullet point expresses the fact that unitary evolution \emph{typically} leads to thermal behavior, with an infinite temperature density matrix for this specific case. The second point expresses the deviations from thermality, showing that the error size when $A\simeq \mathcal{O}(1)$ is exponentially suppressed. This result has been used repeatedly in the past to argue against the Hawking information paradox \cite{hawking}, since the available computations of correlation functions during black hole evaporation are insensitive to the \emph{expected corrections}, of $\mathcal{O}(e^{-S_{\textrm{BH}}})$ if we trust the previous random approximations.

We remark here, see \cite{vijay, harlow} for recent reviews and references therein, that the intuition coming from the random unitaries/states framework has had a strong impact in the context of black physics and quantum thermalization, given a plethora of new concepts, problems and solutions. But we believe that the framework deserves further development, since its present status has also deficiencies. Firstly, the random unitary framework has not been formulated generically in the context of QFT. Secondly, the random unitary evolution defined in \cite{Lloyd} does not conserve any charge, such as energy, momentum, electric charge, etc. Besides, the framework just provides a Gibbs distribution with infinite temperature. It is then interesting to ask the following questions:
\begin{itemize}
\item Can the framework be generalized systematically so as to include conservation of any given charge?
\item Can it be formalized in a QFT context?
\item Can non-infinite effective temperatures emerge in the framework?
\item Do the deviations from thermality differ from the previous case?
\end{itemize}

In this article we develop a framework to solve these questions, and which has interesting different potential applications. We summarize the results here\footnote{Similar approaches to the one developed in this article have been explored in \cite{srednickisub,vijaytypical,nima,barbon,vijaymatrices,vijayent}.}.

\subsection{Summary of results}

The natural answer for the first question is not to choose the Haar ensemble of random  unitaries/states, but to choose the appropriate ensemble of unitary matrices conserving any given set of charges. The problem with this approach is to construct such an ensemble, and we do not see a systematic way of doing so. In this article we will use the Haar ensemble in an appropriate subspace of the Hilbert space, defined by the given initial charges. This procedure can also be seen as implictly defining an esemble of random unitaries different from the Haar ensemble in the full Hilbert space. Within this framework we will provide generic formulae for the typical reduced density matrix and entanglement entropy of any given subsystem and for any given set of conserved charges. These objects turn out to be written generically in terms of `constrained multiplicities', number of different configurations with a given set of constraints. The results connect in a direct and transparent way modern approaches of quantum information theory, concerning for example the computation of entanglement entropies, with more traditional frameworks of microstate counting.

For the second question we argue that the natural way to generalize the framework to a QFT context is to use the Fock space formalism. Fock space provides an ideal setup to define the appropriate subspaces corresponding to the given set of conserved charges. Besides, it is also an ideal setup to define partitions/subsystems by tracing out different set of particles. One possibility is to integrate out all momentum modes but a given one, as done in \cite{vijayent} for the case of vacuum states. We can also integrate out positively charged particles, left movers of a CFT, etc. Interestingly, the formulas we find in the case of random states are generic for all types of subsystems.

For the third and fourth questions we need to sacrifice generality of the analysis and choose concrete models in which to apply the developed formulas. We choose the simplest but non-trivial cases, in which many things can be computed analytically:  a massless scalar field theory in 1+1 dimensions and its generalization to the case of $N$ scalar fields. For a given initial total energy, we derive the typical density matrix and prove that in the thermodynamic limit it is equal to the thermal density matrix. It turns out that the effective temperature is finite, and depends on the initial energy in the way expected from usual thermal considerations. In this way we are able to derive the equation of state of the QFT directly from random dynamics. The equation of state is seen as the typical macroscopic configuration of an underlying unitary quantum model.

We also compute the deviations from thermality, obtaining an important insight. As commented before, for the so-called Page case, developed in \cite{Lloyd, page}, the typical density matrix is \emph{exactly} the thermal distribution withouth the need of taking the thermodynamic limit, and the deviations of the typical entropy from the entropy of the typical reduced density matrix are exponentially suppressed. This means that deviations from thermality are exponentially supressed, see Appendix~\ref{app1} for a complete discussion of all cases. On the other hand, as we will show, in the QFT context the typical density matrix and thererefore the entanglement entropy of the typical density matrix are not directly equal to their thermal counterparts, the difference being of $\mathcal{O}(1/S)$. Therefore, to compute the leading deviations from thermality we just need to compute the entanglement entropy of the typical density matrix, a feature which will simplify matters in future developments. From other perspective, our results pinpoint the microcanonical ensemble as the right ensemble to use when considering pure state unitary evolution.

At any rate, the important upshot is that the errors expected in a scenario in which some charges are conserved are much bigger than the ones commented before, being of $\mathcal{O}(1/S)$ instead of $\mathcal{O}(e^{-S})$.
The applications of these results to conceptual problems involving black holes is still unclear and deserves further development. At any rate it is interesting that:
\begin{itemize}
\item The corrections affect directly the diagonal entries of the density matrix, and therefore should be seen in observables such as mode occupation numbers. They are not non-perturbative, and should be visible in the first $\mathcal{O}(1/S)$ corrections.
\item For any QFT, deviations from thermality can be computed exactly within our framework, and it would be interesting if this can guide the computations of these corrections directly from black hole geometric considerations.
\end{itemize}

At the end of the article we speculate on potential applications of this framework to the relation between geometry and entanglement \cite{ryu,vijayhole}. In particular, our framework can compute entanglement entropies for any desired subsystem, and it is imaginable that we could get important insights into the emergence of near horizon regions through entanglement in the field theory. In previous literature there has been problems with the `integration out' procedure of high momentum modes, or the so-called `holographic renormalization', when trying to get to the near horizon region of black hole backgrounds, and therefore only the asymptotic region is well understood. It seems that our framework may overcome such difficulties in the specific context of random evolution, and might teach us important lessons about event horizons, stretched horizons and near horizon regions of black holes.

\section{Reduced density matrices in Fock space and multiplicities counting}\label{genericprocedure}

Our initial quantum state will be a pure quantum state within a Fock space, which is the natural basis of QFT and certain spin systems:
\begin{equation}\label{Fock}
\vert i_{a},i_{b},\cdots\rangle=\vert\alpha\rangle\;.
\end{equation}
The notation works as follows. Labels $i_{a},i_{b},\cdots$ denote the number of particles in a given representation. Labels $a,b,\cdots$ specify the representation. For example, in a certain theory we could have $a={p,s,q^{r}}$, where $p$ is the momentum, $s$ the spin and $q^{r}$ some charges labelled by $r$. As usual we assume that the Fock states in (\ref{Fock}) are attached with a total conserved energy $E_{\textrm{T}}$ and total conserved charges $Q_{\textrm{T}}^{r}$ of the free theory, where $r$ label different possible charges. Finally, when there is no need to write all the previous labels we will use greek letters  $\alpha, \beta$ to simplify notation. They are just natural numbers running over the states\footnote{We suppose we have a discrete theory, probably with an infinite, but countable number of states such as in QFT. The examples we work out in the article fall in this class, since they are QFT defined on bounded domains.}. For example, normalization of the basis vector is expressed as $\langle \alpha\vert \beta \rangle =\delta_{\alpha,\beta}$.

We now follow the logic described in the introduction. Consider that the full interacting Hamiltonian of the system $H$ commutes with the total charges, so that $E_{\textrm{T}}$ and $Q_{\textrm{T}}^{r}$ are conserved. If we begin with an initial Fock space state $\vert\Psi_{\textrm{in}}\rangle =\vert\beta\rangle$, with definite values of $E_{\textrm{T}}$ and $Q_{\textrm{T}}^{r}$, the various charge conservation laws force the unitary evolution of the initial state to be of the following form:
\begin{equation}\label{constrainedevo}
\vert \Psi (t)\rangle=U(t)\vert\beta\rangle=\sum_{\alpha=1}^{\alpha=\Omega (E_{\textrm{T}},Q_{\textrm{T}}^{r})} \Psi_{\alpha}(t)\vert \alpha\rangle\;,
\end{equation}
where the sum does not run over all Fock states (\ref{Fock}), but only for those $\vert\alpha\rangle$ with definite values of $E_{\textrm{T}}$ and $Q_{\textrm{T}}^{r}$. By definition there are $\Omega (E_{\textrm{T}},Q_{\textrm{T}}^{r})=e^{S(E_{\textrm{T}},Q_{\textrm{T}}^{r})}$ of those, i.e the microcanonical number of states for a given total energy and total charges.

As explained in the previous section, a useful approximation when considering chaotic and highly entropic sectors of the theory is to consider $U(t\gtrsim t_{\textrm{relax}})\simeq U_{\textrm{random}}$. In this article, with the objective of extending the random unitary framework to systems with conserved charges, we will use Haar randomness in the previous subspace, which is indeed defined by the given set of charges. Up to second order corrections (doubly exponentially suppressed in the entropy of the given sector) this approximation can be defined by the following averaging procedure\footnote{In Appendix~{\ref{app1}} we show how to derive Page's formula for the average entanglement entropy \cite{page} with these simple Gaussian relations.}
\begin{equation}\label{ensemble}
[\Psi_{\alpha}]=0~~~~~~[\Psi^{*}_{\alpha}\Psi_{\beta}]=\Lambda \delta_{\alpha\beta}\;.
\end{equation}
Imposing average normalization of the state we obtain
\begin{equation} 
[\langle\Psi_{\textrm{out}}\vert \Psi_{\textrm{out}}\rangle]=\Lambda \Omega (E_{\textrm{T}},Q_{\textrm{T}}^{r})=1\ .
\end{equation}
We want to remark that the ensemble of random states will always be the same, the previous gaussian ensemble of random states, which can be obtained by an analogue ensemble of random unitaries with random gaussian matrix entries. What will change is the sector of states in which gaussian randomness is considered. The sector of states will be the mathematical object which needs to be analyzed in several ways. From the perspective of the full Fock space, the relations~(\ref{ensemble}) implicitly define the action of an ensemble of random unitaries, preserving any given number of conserved charges, on some state of the Hilbert space.

With (\ref{ensemble}) we can now ask questions about typical properties of the field theoretic state $\vert \Psi_{\textrm{out}}\rangle$. For example, the average of the \emph{global} density matrix is given by
\begin{equation}
[\rho_{\textrm{out}}]=[\vert \Psi_{\textrm{out}}\rangle  \langle \Psi_{\textrm{out}}\vert]=\sum_{\alpha=1}^{\alpha=\Omega (E_{\textrm{T}},Q_{\textrm{T}}^{r})}\frac{1}{\Omega (E_{\textrm{T}},Q_{\textrm{T}}^{r})}\vert\alpha\rangle\langle\alpha\vert\;,
\end{equation}
which is just the microcanonical density matrix for a theory with a generic number of conserved charges, obtained here in a straightforward manner by applying random dynamics in the appropriate subspace. Random dynamics seem to pinpoint the maximally mixed microcanonical state, instead of the canonical ensemble.

The next natural step, given the structure provided by Fock space, is to choose one particle type, say $a={p,s,q^{r}}$, and integrate out all the others. Notice that the case of momentum space entanglement, analyzed in \cite{vijayent} for vacuum states, is going to be seen here as a special case, arising when we consider all types of particles within a specified momentum shell. But otherwise the Fock space formalism provides naturally more possibilities, such as the entanglement between positively charged particles and the rest, particles with a given spin and the rest, left and right movers of a CFT, etc. All possible types of sets and reduced dynamics within the Fock basis of the Hilbert space are constructed by joining the appropriate set of $a,b,\cdots$ conforming the type of subsystem we want to study. We thus begin with the simplest one, given by one particle type $a$. We first write the state in the following form
\begin{equation}
\vert \Psi_{\textrm{out}}\rangle=\sum_{i_{a}, \alpha} \Psi_{i_{a}, \alpha}\vert i_{a},\alpha\rangle\;,
\end{equation}
so that now $\alpha$ represents all other particle types different from $a$, but the sum still runs over the previously specified subspace. The reduced density matrix is
\begin{equation}\label{density}
\rho_{a}=\sum_{\alpha}\langle \alpha\vert  \Psi_{\textrm{out}}\rangle\langle\Psi_{\textrm{out}}\vert \alpha\rangle = \sum_{i_{a}} (\sum_{\alpha}\Psi_{i_{a}, \alpha}\Psi^{*}_{i_{a}, \alpha})\vert i_{a}\rangle\langle i_{a}\vert\;.
\end{equation}
This is directly a diagonal density matrix, without the need of averaging. Indeed, for a subsystem with one particle type $a$, it just happens that for two states $\vert i_a,\alpha\rangle$  and $\vert j_a,\alpha\rangle$ with the same energy and charge, they must obey $i_a=j_a$. It will no longer be true for subsystems with more than one particle type.  Using~(\ref{ensemble}) we can obtain the typical reduced density matrix
\begin{eqnarray}\label{typicaldensity}
[\rho_{a}]&=&\sum_{i_{a}=0}^{i^{\textrm{max}}_{a}}\frac{\Omega (E_{\textrm{T}},Q_{\textrm{T}}^{r},i_{a})}{\Omega (E_{\textrm{T}},Q_{\textrm{T}}^{r})}\vert i_{a}\rangle\langle i_{a}\vert \nonumber \\ &=&\sum_{i_{a}=0}^{i^{\textrm{max}}_{a}}P(E_{\textrm{T}},Q_{\textrm{T}}^{r},i_{a})\vert i_{a}\rangle\langle i_{a}\vert\;,
\end{eqnarray}
where $\Omega (E_{\textrm{T}},Q_{\textrm{T}}^{r},i_{a})$ is the number of states with energies $E_{\textrm{T}}$, charges $Q_{\textrm{T}}^{r}$, and a fixed number $i_{a}$ of particles of type $a$. The maximum value of the number of particles of type $a$ is denoted by ${i^{\textrm{max}}_{a}}$ and needs to be determined case by case. We discuss this in the example presented in the next section.

With the previous density matrix, the average entanglement entropy of a particle of a given type $a$, in the fixed $E_{\textrm{T}}$ and $Q_{\textrm{T}}^{r}$ sector is given by:
\begin{equation}\label{entropy}
[S_{\textrm{E}}(\rho_{a})]=-\sum_{i_{a}=0}^{i^{\textrm{max}}_{a}}P(E_{\textrm{T}},Q_{\textrm{T}}^{r},i_{a})\log P(E_{\textrm{T}},Q_{\textrm{T}}^{r},i_{a})\;.
\end{equation}
Being rigorous this is the entanglement entropy of the average density matrix, usually denoted by $S_{\rm E}([\rho_a])$. The full average entropy is computed in Appendix~\ref{app1}, where we show that the difference between the two is exponentially suppressed in the microcanonical entropy. The entropy of the typical density matrix will be enough for us, since we are not interested in the deviations between the typical entropy and the entropy of the typical density matrix, but instead, on the deviations from \emph{thermality}. In the usual case of \cite{page}, typicality and thermality turn out to be exactly the same, and therefore deviations between thermal entropy and average entanglement entropy are equal to deviations between average entanglement entropy and entanglement entropy of the average (see Appendix~\ref{app1} for the complete discussion). We will show that when conserved charges are taken into account this is not longer true. The leading deviations from thermal entropy are those already accounted for by the entropy of the average density matrix, a feature which simplifies present and further developments. At any rate, if needed, the full computations are described in Appendix~\ref{app1}.

Relations~(\ref{typicaldensity}) and \eqref{entropy} are generic formulas which turn out to be written just in terms of `constrained multiplicities', the number of different states with a given set of constraints. Notice that the result for the entanglement entropy is finite, and no divergences occur, even taking into account that we are dealing with quantum field theories. This is obviously because we are carefully applying conservation of charges. Although the multiplicities might be difficult to compute in general, there might be a class of theories in which they can be computed, and explicit connections with thermal density matrices (in particular generalized Gibbs ensembles) might be established. We do this in the following section for the case of certain scalar field theories. At the same time the generality of~(\ref{typicaldensity}) and~(\ref{entropy}) suggests there might be a generic way to prove that the probabilities $P(E_{\textrm{T}},Q_{\textrm{T}}^{r},i_{a})$ are well approximated by the Generalized Gibbs Ensemble. Also it would be interesting to compute this multiplicities for integrable theories. In those theories Generalized Gibbs Ensembles are expected to govern the dynamics but deviations from them might be bigger. This might aliviate several problems encountered in previous literature, see \cite{integrable}. We leave these interesting paths for future work.

Generalizing the procedure to include any desired subset of particles is straightforward. We first form the set of $a,b,\cdots, c$ in which we are interested, and write $\vert \Psi_{\textrm{out}}\rangle$ as:
\begin{equation}
\vert \Psi_{\textrm{out}}\rangle=\sum_{i_{a}, i_{b},\cdots,i_{c},\alpha} \Psi_{i_{a}, i_b,\cdots,i_c, \alpha}\,\vert i_{a},i_b,\cdots, i_c;\alpha\rangle\;,
\end{equation}
where $\alpha$ now labels all particles types different from $a,b,\cdots, c$. The reduced density matrix then reads:
\begin{eqnarray}\label{reducedsubset}
\rho_{a,b,\cdots ,c}&=&\sum_{\alpha}\langle \alpha\vert  \Psi_{\textrm{out}}\rangle\langle\Psi_{\textrm{out}}\vert \alpha\rangle \\
&=& \sum_{i,i'} \left(\sum_{\alpha}\Psi_{i_{a},i_{b}, \cdots ,i_{c},\alpha}\Psi^{*}_{i'_{a}, i'_{b},\cdots , i'_{c}, \alpha}\right)\vert i_{a},i_{b},\cdots ,i_{c}\rangle\langle i'_{a},i'_{b},\cdots ,i'_{c}\vert\ ,\nonumber
\end{eqnarray}
where $\alpha$ labels the set of particles which are traced out. As opposed to the previous one-particle case, this is not a diagonal density matrix\footnote{Although it is not diagonal, it has a nice block diagonal structure, as used and described in Appendix~\ref{app1}.}. However, taking the average we find the following generic formula:
\begin{eqnarray}\label{typicalreducedsubset}
[\rho_{a,b,\cdots ,c}]&=&\sum_{i}\frac{\Omega (E_{\textrm{T}},Q_{\textrm{T}}^{r},i_{a},i_{b},\cdots ,i_{c})}{\Omega (E_{\textrm{T}},Q_{\textrm{T}}^{r})}\vert i_{a},i_{b},\cdots ,i_{c}\rangle\langle i_{a},i_{b},\cdots ,i_{c}\vert =\nonumber \\ &=&\sum_{i}P(E_{\textrm{T}},Q_{\textrm{T}}^{r},i_{a},i_{b},\cdots ,i_{c})\vert i_{a},i_{b},\cdots ,i_{c}\rangle\langle i_{a},i_{b},\cdots ,i_{c}\vert\;,
\end{eqnarray}
which is diagonal. The average probabilities $P(E_{\textrm{T}},Q_{\textrm{T}}^{r},i'_{a},i'_{b},\cdots ,i'_{c})$ are ready to be compared with thermal expectations, i.e with probabilities coming from Gibbs distributions. We will discuss examples in the next section.

The average entanglement entropy - i.e. the entanglement entropy of the average density matrix (see Appendix \ref{app1})- is finally
\begin{equation}\label{entanglementsubset}
[S_{\textrm{E}}(\rho_{a,b,\cdots , c})]=-\sum_{i}P(E_{\textrm{T}},Q_{\textrm{T}}^{r},i_{a},i_{b},\cdots ,i_{c})\log P(E_{\textrm{T}},Q_{\textrm{T}}^{r},i_{a},i_{b},\cdots ,i_{c})\;.
\end{equation}
We concldue that the generic equations \eqref{typicalreducedsubset} and \eqref{entanglementsubset} apply to any theory with a Fock basis structure, such as a QFT or certain spin systems. In particular it is applicable to Holographic Field theories. It seems a rigorous framework to study entanglement between infrared and ultraviolet domains at finite temperature. An exciting possibility is that we could potentially extract physics from the near horizon regions of quantum black holes from the structure of entanglement in Fock space of the field theory. At any rate it is expected to give more insights into the connections between entanglement and quantum gravity \cite{ryu,vijayhole}, an exciting direction which we leave for future work.

%%%%%%%%%%%%%%%%%%%%%%%%%%%%%%%%%%%%%%%%
\section{Examples: massless scalar fields in two dimensions}
%%%%%%%%%%%%%%%%%%%%%%%%%%%%%%%%%%%%%%%%

In this section we analyze in detail two specific examples. The first example is a massless scalar field theory in 1+1 dimensions defined on a finite line segment. The second one is its generalization to the case of $N$ scalar fields. For both cases we will show the emergence of Gibbs distributions as effective states for reduced subsystems. The procedure also allows the computation of deviations from precise thermality. Finally, entanglement entropy for single modes as a function of the momentum, and entanglement between high and low energy momentum modes for a given energy cutoff will be discussed. We will end with the analogue of Page curve \cite{page} for the case at hand. This enlarges the program spelled out in \cite{vijayent} to the case of random states.

\subsection{Massless real scalar field on a finite segment}

In this case the energy/momentum dispersion relation for excitations over the vacuum, together with the quantization condition reads
\begin{equation}
E_{n}=p_{n}=\frac{\pi n}{L}\;,
\end{equation}
for a segment of length $L$ and $n=1,2,\cdots$. The Fock space is spanned by vectors of the type:
\begin{equation}\label{fockstates}
\vert \alpha\rangle\equiv\vert i_{n=1},i_{n=2},\cdots\rangle = \vert i_{1},i_{2},\cdots\rangle\;,
\end{equation}
where $i_{n}$ is the number of particles with momentum $p_{n}$, and $\alpha$ is just a natural number running over all the infinite but countable eigenstates, used here to simplify notation. These are eigenstates of the free Hamiltonian with Dirichlet boundary conditions. If the true Hamiltonian conserves the total energy, which for a Fock state reads
\begin{equation}\label{energyscalar}
E_{\textrm{T}}=\sum_{n=1}^{n=n_{\textrm{max}}}i_{n} p_{n} = \frac{\pi}{L}\sum_{n=1}^{n=n_{\textrm{max}}}i_{n} n\;,
\end{equation}
where $n_{\textrm{max}}$ is defined by $p_{n_{\textrm{max}}}=E_{\textrm{T}}$, then any initial state with definite total energy $E_{\textrm{T}}$ will evolve towards states of the form
\begin{equation}
\vert\Psi_{\textrm{out}}\rangle=\sum_{\alpha=1}^{\alpha=\Omega (E_{\textrm{T}})}\Psi_{\alpha}\vert \alpha\rangle\;,
\end{equation}
where the $\vert \alpha\rangle$ are all the states belonging to (\ref{fockstates}) with total energy $E_{\textrm{T}}$.  By definition these are
$\Omega (E_{\textrm{T}})=e^{S (E_{\textrm{T}})}$, where $S (E_{\textrm{T}})$ is the microcanonical entropy at energy $E_{\textrm{T}}$. Noticing that (\ref{energyscalar}) can be written as
\begin{equation}\label{partition}
\frac{L}{\pi}E_{\textrm{T}}=\sum_{n=1}^{n=n_{\textrm{max}}} i_{n} n\;,
\end{equation}
we conclude that the number of states with a given $E_{\textrm{T}}$ is equal to the number of different partitions $p(n)$ of the natural number $n_{\textrm{max}}=\frac{L}{\pi}E_{\textrm{T}}$, which is given asymptotically for large $n_{\textrm{max}}$ by
\begin{equation}\label{asymptotics}
\Omega (E_{\textrm{T}})=p(\frac{L}{ \pi}E_{\textrm{T}})\rightarrow \frac{\pi}{4 \sqrt{3} E_{\textrm{T}} L}e^{\sqrt{\frac{2\pi E_{\textrm{T}}L}{3}}}\;.
\end{equation}
The global entropy can now easily be extracted and for large  $n_{\textrm{max}}$ we find
\begin{equation}\label{totalentropy}
S={\sqrt{\frac{2\pi}{3} L E_{\textrm{T}}}}=2\pi{\sqrt{\frac{1}{6}\, n_{\textrm{max}}}}\ .
\end{equation}
So, in the limit $E_{\rm T}\rightarrow \infty$, this agrees with the Cardy formula for a CFT with central charge $c=1$ and $L_0-\frac{c}{24}=n_{\textrm{max}}$. If the interacting theory would be a conformal theory on a line segment of length $L$, the conformal dimension of the pure state $\vert\Psi_{\textrm{out}}\rangle$ would then be $\Delta = n_{\rm max}$.

Now we can directly apply the generic formulas derived in the previous section. The approximation $U(t\gtrsim t_{\textrm{relax}})\simeq U_{\textrm{random}}$ is operationally defined by (\ref{ensemble}), which we repeat here to emphasize that it does not change from one case to another:
\begin{equation}
[\Psi_{\alpha}]=0~~~~~~[\Psi^{*}_{\alpha}\Psi_{\beta}]=\Lambda \delta_{\alpha,\beta}\;.
\end{equation}
Imposing average normalization, we obtain $[\langle\Psi_{\textrm{out}}\vert \Psi_{\textrm{out}}\rangle]=\Lambda \Omega (E_{\textrm{T}})=1 $. The average of the global density matrix is given by
\begin{equation}
[\rho_{\textrm{out}}]=[\vert \Psi_{\textrm{out}}\rangle  \langle \Psi_{\textrm{out}}\vert]=\sum_{\alpha=1}^{\alpha=\Omega (E_{\textrm{T}})}\frac{1}{\Omega (E_{\textrm{T}})}\vert\alpha\rangle\langle\alpha\vert\;,
\end{equation}
which is just the microcanonical density matrix, obtained here in a straightforward manner by applying random dynamics in the appropriate subspace.

\subsubsection{Entanglement of a single momentum cell}

We now study the entanglement entropy in a pure state of a single particle specie.
In our example, choosing one particle type just amounts to choosing a definite momentum $p_{n}$ for a fixed $n$. The momentum cell can still be multiply occupied, so the subsystem has a finite dimension determined by the size of the cell, i.e. $i_n^{\rm max}+1$, where the maximum occupation number is given by $i^{\textrm{max}}_{n}=\lfloor E_{\textrm{T}}/p_{n}\rfloor$, the integer closest to $ E_{\textrm{T}}/p_{n}$ from below. Writing the state in the form
\begin{equation}
\vert \Psi_{\textrm{out}}\rangle=\sum_{i_{n}, \alpha} \Psi_{i_{n}, \alpha}\vert i_{n},\alpha\rangle\;,
\end{equation}
where $\alpha$ now represents all other particles types different from $p_{n}$, we arrive at
\begin{equation}
\rho_{n}=\sum_{\alpha}\langle \alpha\vert  \Psi_{\textrm{out}}\rangle\langle\Psi_{\textrm{out}}\vert \alpha\rangle = \sum_{i_{n}=0}^{i^{\textrm{max}}_{n}} (\sum_{\alpha}\Psi_{i_{n}, \alpha}\Psi^{*}_{i_{n}, \alpha})\vert i_{n}\rangle\langle i_{n}\vert\;.
\end{equation}
This is directly a diagonal density matrix, due to energy conservation. The average density matrix is
\begin{equation}\label{density2}
[\rho_{n}]=\sum_{i_{n}=0}^{i^{\textrm{max}}_{n}}\frac{\Omega (E_{\textrm{T}},i_{n})}{\Omega (E_{\textrm{T}})}\vert i_{n}\rangle\langle i_{n}\vert=\sum_{i_{n}=0}^{i^{\textrm{max}}_{n}}P(E_{\textrm{T}},i_{n})\vert i_{n}\rangle\langle i_{n}\vert\;,
\end{equation}
where $\Omega (E_{\textrm{T}},i_{n})$ is the number of states with total energy $E_{\textrm{T}}$ and $i_{n}$ particles with momentum $p_{n}$. Therefore, the average entanglement entropy of  momentum cell $n$ is given by
\begin{equation}\label{entropy2}
[S_{\textrm{E}}(\rho_{n})]=-\sum_{i_{n}=0}^{i^{\textrm{max}}_{n}}P(E_{\textrm{T}},i_{n})\log P(E_{\textrm{T}},i_{n})\;.
\end{equation}
Formulas \eqref{density2} and \eqref{entropy2} are the analogues of \eqref{typicaldensity} and\eqref{entropy} for the case at hand. To compute these quantitities we need to find $\Omega (E_{\textrm{T}},i_{n})$. In this example this number can be analytically computed. The procedure is explained in detail in Appendix~\ref{app2}. Here we just quote the result:
\begin{equation}
\Omega (E_{\textrm{T}},i_{n})=p\left(\frac{L}{ \pi}E_{\textrm{T}}-i_{n} n\right)-p\left(\frac{L}{ \pi}E_{\textrm{T}}-n (i_{n} +1)\right)\;,
\end{equation}
where $p(n)$ is the number of partitions of the number $n$. 

Now we are ready to make the first crosscheck of our computations, which constitutes one of the main results of the article. The typical probability of finding $i_{n}$ particles with momentum $p_{n}$ is given by 
\begin{equation}\label{prob-one-cell}
P(E_{\textrm{T}},i_{n})=\frac{\Omega (E_{\textrm{T}},i_{n})}{\Omega (E_{\textrm{T}})}=\frac{p(\frac{L}{ \pi}E_{\textrm{T}}-i_{n} n)-p(\frac{L}{ \pi}E_{\textrm{T}}-n (i_{n} +1))}{p(\frac{L}{ \pi}E_{\textrm{T}})}\;.
\end{equation}
It is easy to check that all probabilities add up to unity. The leading term when $i_{n}p_{n}\ll E_{\textrm{T}}$, or equivalently $i_n\ll i_n^{\rm max}$, is given by
\begin{equation}
P(E_{\textrm{T}},i_{n})\rightarrow e^{-\sqrt{\frac{L\pi}{6 E_{\textrm{T}}}}i_{n} p_{n}}-
e^{-\sqrt{\frac{L\pi}{6 E_{\textrm{T}}}} p_{n}(i_{n}+1)}\:.
\end{equation}
This is a thermodynamic limit. Indeed, $i_{n}p_{n}\ll E_{\textrm{T}}$ means that the energy in momentum cell $n$ is small compared to the total energy, or equivalently, momentum cell $n$ is not well occupied. This implies that most of the energy is sitting in the other momentum cells, and that these cells form a heat bath for momentum cell $n$. Another way of saying this, is that this limit is a good approximation for low values of $n$. For higher values of $n$, the energy in this momentum cell will typically be too large, or $i_n^{\rm max}=\lfloor n_{\rm max}/n \rfloor$ will be too small for the thermodynamic limit to be a good approximation. If we would have used the usual Gibbs ensemble this probability would be given by 
\begin{equation}\label{Gibbs-prob}
P(T,i_{n})=\frac{e^{-\beta i_{n}p_{n}}}{\sum_{m=0}^{\infty}e^{-\beta m p_{n}}}=e^{-\beta i_{n} p_{n}}-
e^{-\beta p_{n}(i_{n}+1)}\;.
\end{equation}
So in the limit $i_{n}p_{n}\ll E_{\textrm{T}}$, of for CFT's, large conformal dimensions $\Delta=n_{\rm max}$, the pure state is typically seen by the momentum cell as a thermal bath at a temperature given by
\begin{equation}\label{Gibbs-temp}
T=\sqrt{\frac{6 E_{\textrm{T}}}{\pi L}}\;.
\end{equation}
We thus see that it is possible to derive generic Gibbs ensembles, with any desired effective temperature, using random dynamics in the approppriate subspace. Notice that \eqref{Gibbs-temp} implies $S(E_{\textrm{T}})T(E_{\textrm{T}})=2E_{\textrm{T}}$ on the one hand, and on the other hand there is the general relation $S(E_{\textrm{T}})T(E_{\textrm{T}})=E_{\textrm{T}}+{\cal {P}}V$. Combining the two, we can determine the pressure density: 
\begin{equation}
{\cal{P}}=E_{\textrm{T}}/V=\epsilon\;.
\end{equation}
This is the same equation of state as for a two-dimensional CFT. In this way we expect to recover the known relation $\epsilon=(d-1){\cal P}$, valid for $d$-dimensional CFT's, using just random dynamics. More generically it should be possible to derive any equation of state by choosing the appropriate field theory and using the same procedure. The equation of state of a fluid system, at least for this case, is the typical macroscopic configuration of the true evolving pure quantum state.

The next step is to compute the deviations form thermality. The next to leading term in the previous $i_{n}p_{n}\ll E_{\textrm{T}}$ expansion is given by
\begin{equation}\label{error}
P(E_{\textrm{T}},i_{n})=P(T,i_{n})+\frac{i_{n} p_{n}}{E_{\textrm{T}}}e^{-\beta i_{n} p_{n}}-\frac{p_{n}(i_{n}+1)}{E_{\textrm{T}}}e^{-\beta p_{n}(i_{n}+1)}=P(T,i_{n})+\textrm{error}\;.
\end{equation}
To proof this, we only needed the Hardy-Ramanujan asymptotic formula for the number of partitions, see \eqref{prob-error} in Appendix \ref{app2}.

The error we find here is thus much bigger than the one is usually expected. Within the usual random unitary formalism, the errors are typically of $\mathcal{O}(e^{-S})$, so exponentially suppressed in the entropy. In our case, if we take a momentum cell with energy $i_np_n\simeq T$, we are finding errors of $\mathcal{O}(T/E)\sim\mathcal{O}(1/S)$. This seems just to be due to energy conservation, which is explicitly ensured in our formalism. That we find such large errors may have implications for black hole physics and bulk locality in AdS/CFT, see \cite{vijay,harlow} and references therein, and we address this issue further in the next subsection when we discuss the large $N$ limit.
 
Finally, the average entanglement entropy of the momentum mode is given by\footnote{In the sum it might happen that some probabilities are zero. These terms simply do not contribute to the sum, since $\lim_{x\rightarrow 0}x\log x=0$.} :
\begin{equation}\label{ent-ent}
[S_{\textrm{E}}(n,n_{\textrm{max}})]=-\sum_{i_{n}=0}^{i_{n}^{\textrm{max}}}P(E_{\textrm{T}},i_{n})\log P(E_{\textrm{T}},i_{n})\ ,
\end{equation}
and we remind that $i_n^{\textrm{max}}=\lfloor E_{\textrm{T}}/p_n\rfloor=\lfloor n_{\textrm{max}}/n \rfloor$. The dominant contributions to this sum come from the low occupation numbers, where the probabilities are approximately thermal. For high occupation numbers, $i_n\simeq i_n^{\rm max}$, the probabilities are exponentially suppressed in the entropy, as one can easily compute from \eqref{prob-one-cell}.

At present we have not found a way to evaluate the sum algebraically in closed form as a function of $n$, though one can evaluate the sums explicitly for any given $n$ and $n_{\textrm{max}}$, e.g. on Mathematica. However, for high values of $n$ this is possible. For example, in the case of $n=n_{\textrm{max}}$, the highest momentum possible, there are only two terms in the sum in the entanglement entropy, $i_{n}=0$ and $i_{n}=1=i_n^{\rm max}$, and the result is
\begin{eqnarray}\label{entropyhigh}
[S_{\textrm{E}}(n_{\rm max}, n_{\textrm{max}})]&=&-\left(1-\frac{1}{p(n_{\textrm{max}})}\right) \log\left(1-\frac{1}{p(n_{\textrm{max}})}\right)-\frac{1}{p(n_{\textrm{max}})} \log\frac{1}{p(n_{\textrm{max}})}\nonumber\\
&\approx& S\,e^{-S}\:,
\end{eqnarray}
where in the second line, we took the leading term in the limit $n=n_{\textrm{max}}\rightarrow \infty$. The relation~(\ref{entropyhigh}) shows that there are extremely small entanglement entropies in random QFT states\footnote{Notice that this is not the case when considering the common Page case \cite{page}. Indeed, in that case the minimum entanglement entropies are of $\mathcal{O}(1)$ in the thermodynamic limit.}. For conformal field theories, and in the context of the AdS/CFT correspondence, we do not expect these entanglement entropies to be captured by some geometric quantities in the bulk. The proposal for deriving entanglement entropy of CFT's holographically, developed in \cite{ryu}, is expected to capture entanglement entropies with a minimum size of $\mathcal{O}(1)$, since this would correspond to surfaces of Planckian size. Entanglement entropies of $\mathcal{O}(Se^{-S})$ are clearly of non-perturbative nature from the point of view of AdS/CFT dualities \cite{maldacena}.

On the other hand, the thermal entropy of a single mode is based on the Gibbs probability distribution \eqref{Gibbs-prob}. The result is well-known and can be directly computed using $P(E_{\textrm{T}},i_{n})$ and Shannon's expression for the entropy. It reads
\begin{equation}\label{thermaln}
S_\beta(n)=\beta p_n\, \frac{e^{-\beta p_n}}{1-e^{-\beta p_n}}-\log(1-e^{-\beta p_n})\ ,
\end{equation}
with inverse temperature $\beta=1/T$ given by \eqref{Gibbs-temp}, and we remind that $p_n=\pi n/L$, such that $\beta p_n=n\pi /{\sqrt{ 6n_{\textrm{max}}}}$, which is independent of the size $L$. The  function \eqref{thermaln} is a monotonically decaying function.

\begin{figure}[h]
\centering
{\includegraphics[height=42mm]{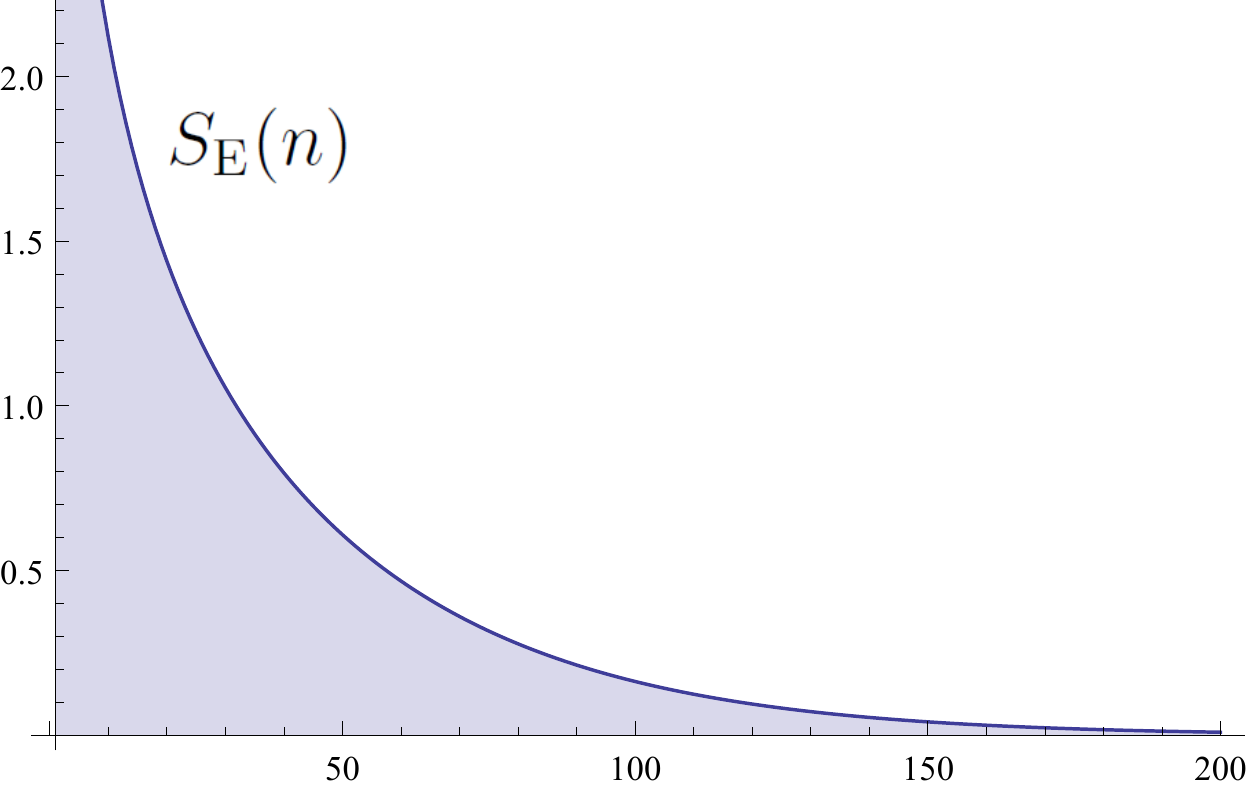}}
{\includegraphics[height=42mm]{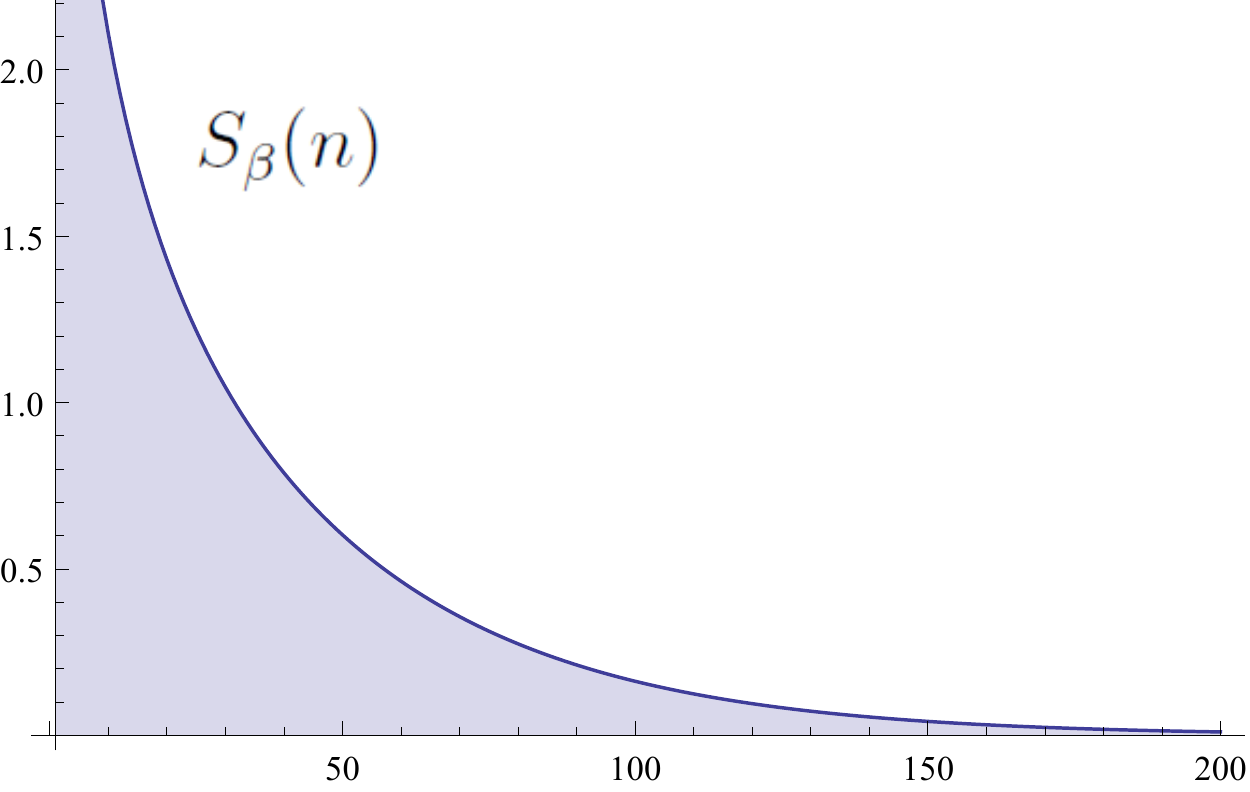}}
\caption{{\footnotesize Average entanglement entropy $S_{\textrm{E}}(n)$ (left) of a single momentum cell $n$ and thermal entropy $S_\beta(n)$ (right) seen by the momentum cell, for $n_{\textrm{max}}=1500$ and $n$ varying between 1 and 200 (horizontal axis). For this large value of $n_{\textrm{max}}$, the agreement is very good.}}
\label{fig1}
\end{figure}

The two functions, the entanglement entropy and the thermal entropy, are plotted against the momentum $n$ in Figure \ref{fig1}. The dependence against $n$ provides some short of entanglement `running' on the energy scale of a thermal-like state.

\begin{figure}[h]
\centering
{\includegraphics[height=42mm]{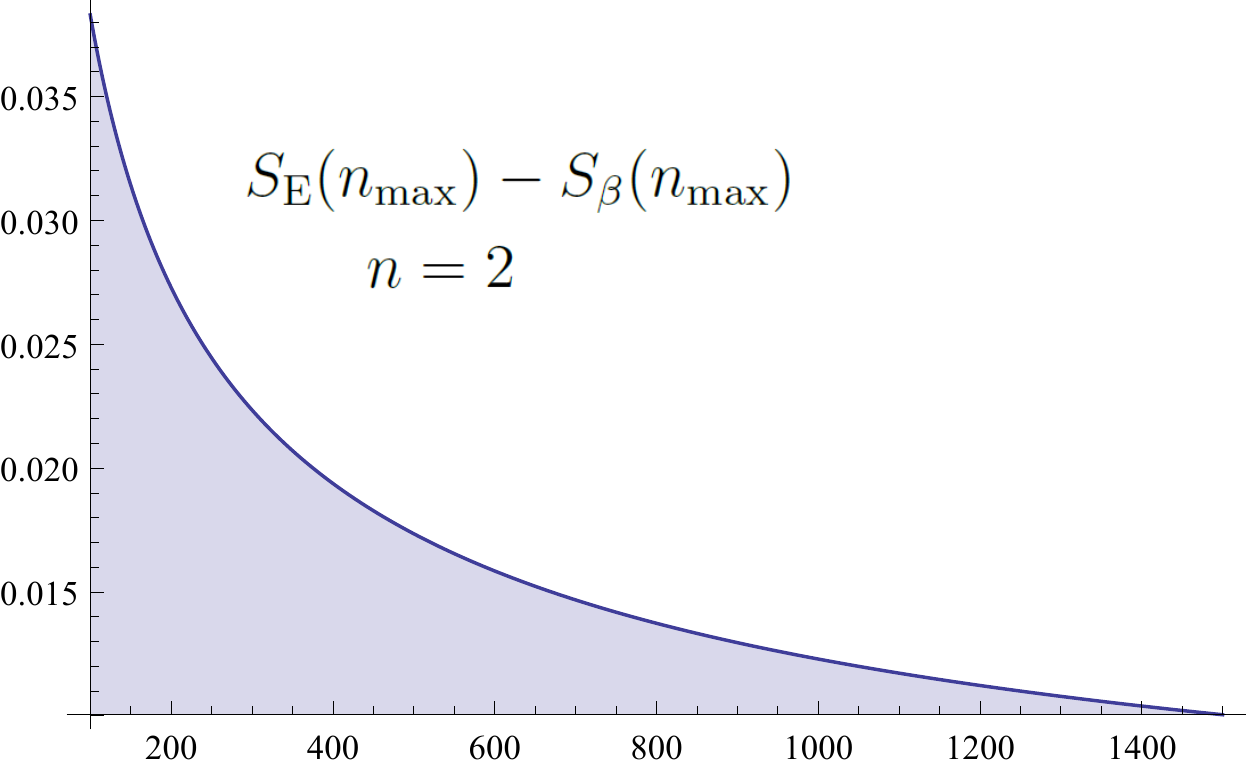}}
\caption{{\footnotesize Difference of entanglement and thermal entropy, now as a function of $n_{\textrm{max}}$, and for fixed $n$ here taken to be $n=2$. A best fit shows that the fall-off of this graph goes approximately like $1/{\sqrt {n_{\textrm{max}}}}\approx 1/S$, and not like $e^{-S}$. As explained earlier in the text, this only holds for low values of $n$. For large values of $n$, the differences become larger.}}
\label{fig2}
\end{figure}

It would be interesting to find an asymptotic formula for the entanglement entropy for large  $E_{\textrm{T}}$ and generic momentum $p_{n}$. This would make easier the comparison with the thermal approximation. Numerical analysis, as can be seen from Figure \ref{fig2}, shows that for large values of $n_{\textrm{max}}$, the entanglement entropy approaches the thermal entropy, but the corrections to the thermal result are not exponentially suppressed in the entropy, and go like $1/{\sqrt{n_{\rm{max}}}}\approx 1/S$. This was obviously expected from the corresponding errors in the probabilities themselves. At any rate, notice that since corrections die in the thermodynamic limit, the expectations coming from the simple analytical thermal expression~(\ref{thermaln}) might be ready to compare with geometric dual formulations of the QFT, a very interestring direction to explore.

\subsubsection{Entanglement entropy of low-energy degrees of freedom and Page curve}

\begin{figure}[h]
\centering
{\includegraphics[height=42mm]{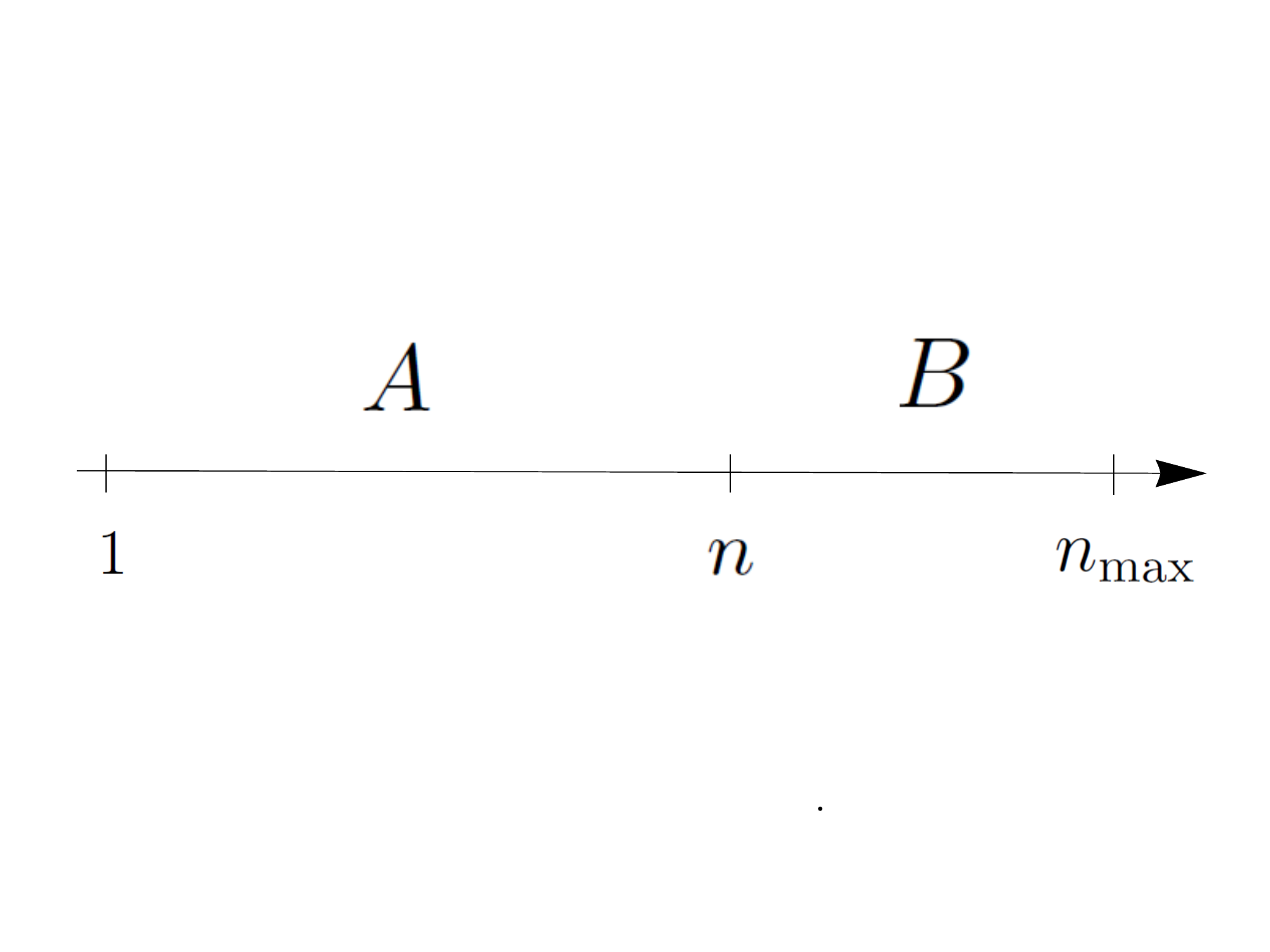}}
\caption{{\footnotesize Two complementary subsystems $A$ and $B$ in Fock space. The entanglement entropy of these subsystems  is computed below.}}
\label{fig4}
\end{figure}

As described before we can also consider any subset of particles we wish. In this case a natural property to analyze is the entanglement between high-momentum and low-momentum modes for a given threshold $p_{n}$. More explicitly we can integrate out all particles with momentum higher than $p_{n}$. The reduced density matrix is then given by
\begin{equation}\label{reducedscalar}
\rho_{1,2,\cdots,n}=\sum_{i,i'} (\sum_{\alpha}\Psi_{i_{1},i_{2}, \cdots ,i_{n},\alpha}\Psi^{*}_{i'_{1},i'_{2},\cdots ,i'_{n} \alpha})\vert i_{1},i_{2}, \cdots ,i_{n}\rangle\langle i'_{1},i'_{2},\cdots ,i'_{n}\vert\;,
\end{equation}
where $\alpha$ labels the set of particles with momentum greater than $p_{n}$. As in the previous section, this is not a diagonal density matrix. Taking the average we find
\begin{eqnarray}\label{typicalreducedscalar}
[\rho_{1,2,\cdots ,n}]&=&\sum_{i}\frac{\Omega (E_{\textrm{T}},i_{1},i_{2}, \cdots ,i_{n})}{\Omega (E_{\textrm{T}})}\vert i_{1},i_{2}, \cdots ,i_{n}\rangle\langle i_{1},i_{2},\cdots ,i_{n}\vert =\nonumber \\ &=&\sum_{i,i}P(E_{\textrm{T}},i_{1},i_{2}, \cdots ,i_{n})\vert i_{1},i_{2}, \cdots ,i_{n}\rangle\langle i_{1},i_{2},\cdots ,i_{n}\vert\;.
\end{eqnarray}
The average probabilities $P(E_{\textrm{T}},i_{1},i_{2},\cdots ,i_{n})$ can be computed using the same previous method of constrained partitions. This is explained in detail in Appendix~\ref{app2}. The result is given by~(\ref{lownumber}) divided by the total number of partitions $p(\frac{L}{2 \pi}E_{\textrm{T}})$:
\begin{eqnarray}\label{probn}
P (E_{\textrm{T}},i_{1},i_{2},\cdots ,i_{n})&=& \frac{1}{\Omega (E_{\textrm{T}})}\Biggl\{ p\left(\frac{L}{2 \pi}E_{\textrm{T}}-\sum_{k=1}^{n}i_{k} k\right)+ \nonumber\\
&-&\sum_{l=1}^{n}p\left(\frac{L}{2 \pi}E_{\textrm{T}}-\sum_{k=1}^n i_{k} k-l\right)+ \nonumber \\
&+&\sum_{l<r}p\left(\frac{L}{2 \pi}E_{\textrm{T}}-\sum_{k=1}^n i_{k} k-l-r\right)+\cdots \nonumber\\
&+&(-1)^{n}p\left(\frac{L}{2 \pi}E_{\textrm{T}}-\sum_{k=1}^n i_{k} k-\sum_{l=1}^n l\right)\Biggr\} \ .
\end{eqnarray}
This is a quite complicated expression for generic $i_{1},i_{2}, \cdots ,i_{n}$, though one can evaluate these sums using Mathematica rather easily. The situation becomes more cumbersome when evaluating the entanglement entropy, since we now have to perform additional sums over multiple occupation numbers. In practice this turns out to be rather hard using Mathematica, and we only succeeded to evaluate the entanglement entropy of a system of up to the first eight modes. However, a simplification occurs whenever $\sum_{k}^n i_{k} p_k\ll E_{\textrm{T}}$, namely when the subsystem has much smaller energy then the total energy. Notice that if $r\ll n$, then we can approximate the partitions
\begin{equation}\label{approx-partition}
p(n-r)\simeq p(n)e^{-\frac{\pi r}{\sqrt{6 n}}}\;,
\end{equation}
which we can use in the probabilities \eqref{probn}. Indeed, if $\sum_{k=1}^n i_{k} p_k\ll E_{\textrm{T}}$ holds, we can apply \eqref{approx-partition} to all terms in~(\ref{probn}), so that we obtain
 \begin{eqnarray}\label{multi-prob}
 P(E_{\textrm{T}},i_{1},i_{2},\cdots ,i_{n})&\simeq & \frac{e^{-\beta \sum_{k=1}^{n}i_{k}p_{k}}}
{\prod_{l=1}^n \left(\sum_{m=0}^\infty e^{-\beta m p_{l}}\right)}\\
&=&e^{-\beta \sum_{k=1}^{n}i_{k}p_{k}} \prod_{l=1}^{n} (1-e^{-\beta p_{l}})=P(T,i_{1},i_{2},\cdots ,i_{n})\;,\nonumber
 \end{eqnarray}
where $P(T,i_{1},i_{2},\cdots ,i_{n})$ is the probability given by the Gibbs ensemble. One can again compute the subleading corrections, using \eqref{prob-expand} for each term in \eqref{multi-prob} and find power-law suppressed terms in the energy of the subsystem divided by the total energy. So the error analysis is similar to the case of a single momentum cell, and the leading corrections scale like $1/S$. Hence for any subsystem, up to computable corrections, the entanglement entropy associated to $\rho_{1,2,\cdots,n}$ is just the sum of thermal entropies of each momentum mode. In this way we can derive the analogous curve in a QFT setup to the so-called Page's curve \cite{page}. Obviously, if we express the entanglement entropy as a function of the corresponding thermal entropy we obtain directly Page's curve, with somewhat different deviations from thermality. But in this context it is more interesting to paint the entanglement entropy as a function of the energy scale used to divide the high-energy modes from the low-energy ones, given by $p_{n}$.
Hence we define two complementary subsystems $A$ and $B$ in momentum space, as depicted in Figure \ref{fig4}.

\begin{figure}[h]
\centering
{\includegraphics[height=42mm]{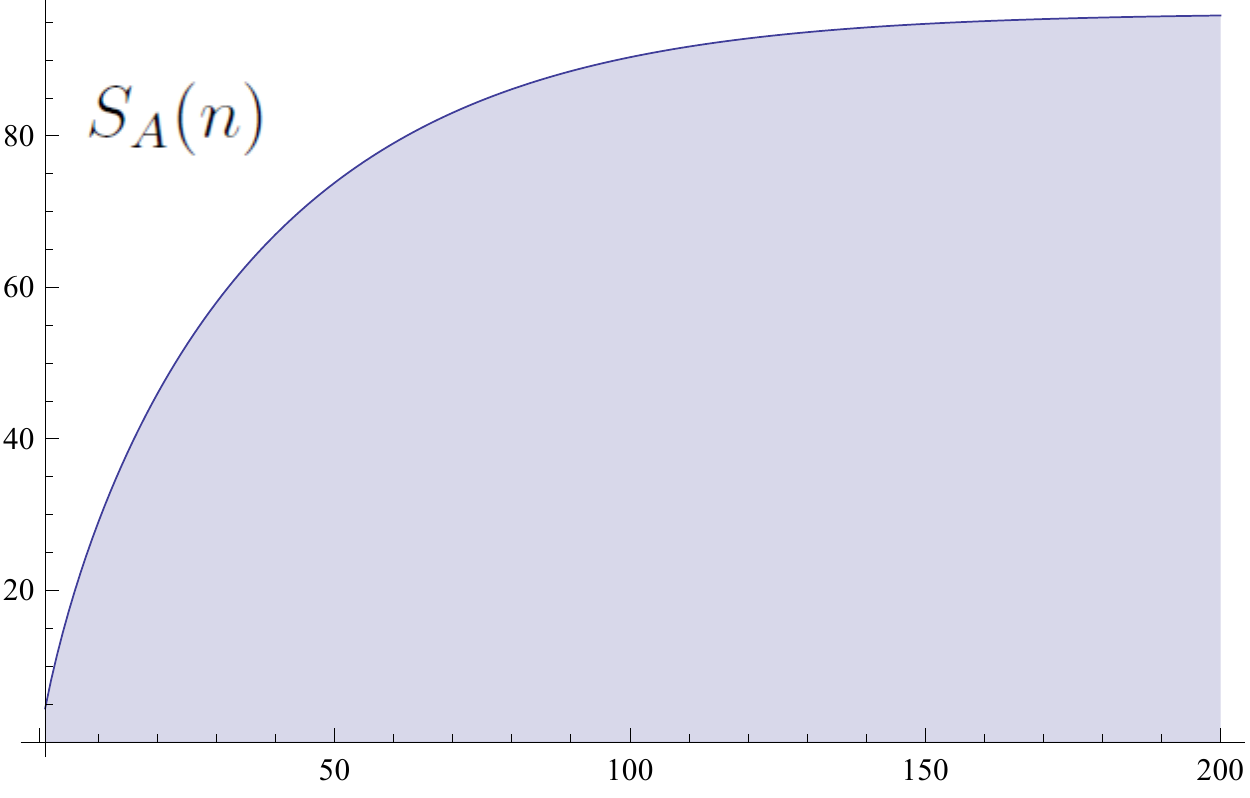}}
\caption{{\footnotesize Thermal entropy (vertical axis) of a subsystem $A$ containing all momentum modes up to a given $n$, as a function of $n$ (horizontal axis). In the figure we took $1\leq n \leq 200$ and $n_{\rm{max}}=1500$. The curve asymptotes, for $n\rightarrow n_{\rm{max}}$, to 96.22, which is reasonably close to the total entropy of the system given by \eqref{totalentropy}, which for  $n_{\rm{max}}=1500$ yields $S\simeq 99.35$. The discrepancy gets smaller for larger $n_{\rm{max}}$. The entanglement entropy approximates the thermal entropy well whenever the energy in subsystem $A$ is small compared to the total energy, so for small values of $n$. Deviations then go like $1/S$.}}
\label{fig3}
\end{figure}

\begin{figure}[h]
\centering
{\includegraphics[height=42mm]{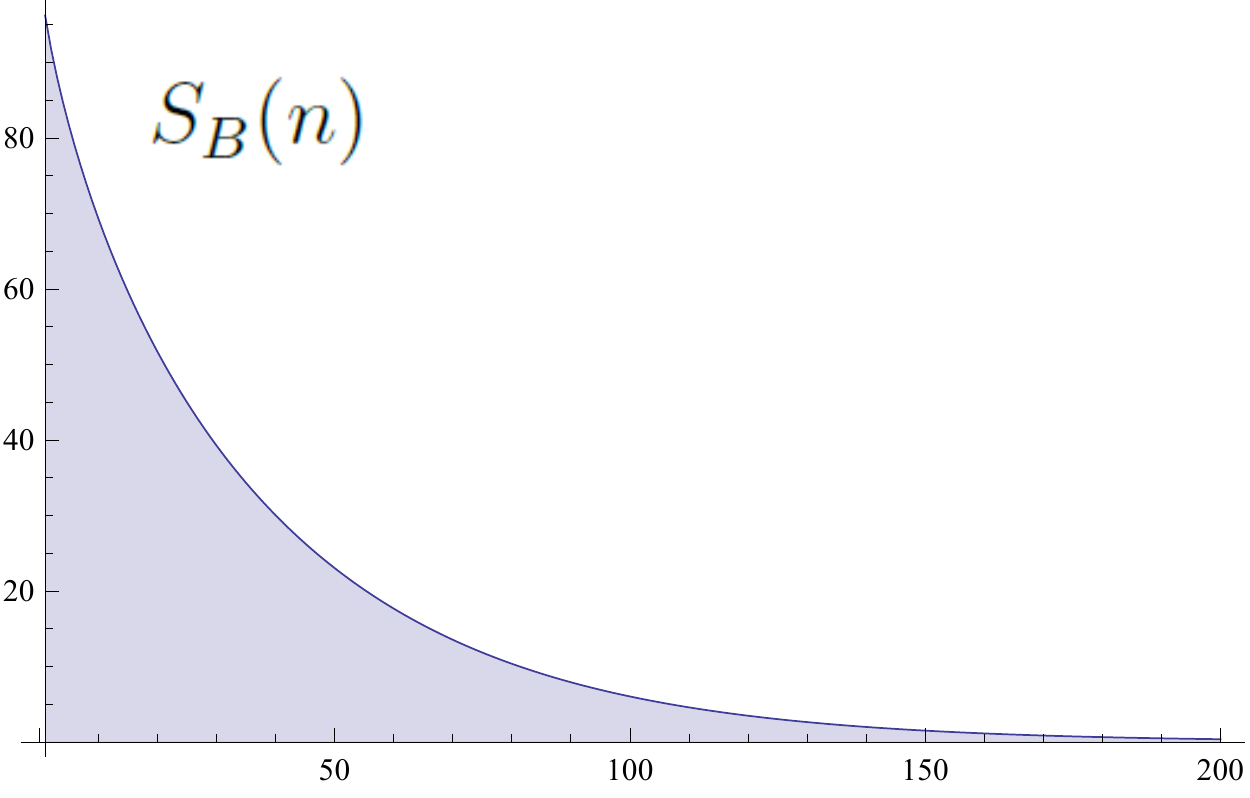}}
\caption{{\footnotesize Thermal entropy $S_B(n)$  (vertical axis) of the complementary subsystem $B$ containing all momentum modes between $n$ and $ n_{\rm{max}}$, as a function of $n$ (horizontal axis). In the figure we took $1\leq n \leq 200$ and $n_{\rm{max}}=1500$, just as in Figure \ref{fig3}. The curve now asymptotes for $n\rightarrow 1$, again to 96.22, which is reasonably close to the total entropy of the system given by \eqref{totalentropy}, which for  $n_{\rm{max}}=1500$ yields $S\simeq 99.35$. The discrepancy gets smaller for larger $n_{\rm{max}}$. The thermal entropy of system $B$ is reduced to half at mode number around $n\approx 21$ again. For larger values $n  \gg 21$, the thermal entropy is a good approximation for the entanglement entropy, with corrections of order $1/S$.}}
\label{fig5}
\end{figure}

 Up to subleading corrections in the thermodynamic limit, we can use the thermal entropy for the low-energy modes, which becomes simply the sum of the thermal entropy of the individual modes
\begin{equation}\label{Wilsonthermalentropy}
S_A (n) \equiv \sum_{k=1}^{n}S_\beta(k)=\sum_{k=1}^{n}\left(\beta p_k\,  \frac{e^{-\beta p_k}}{1-e^{-\beta p_k}}-\log(1-e^{-\beta p_k})\right) \ .
\end{equation}
The plot of the thermal entropy $S_A$ is given in Figure \ref{fig3}.
It is hard to perform the sum in \eqref{Wilsonthermalentropy} analytically, except in the continuum limit where the modes $\beta p_n$ become infinitesimally spaced. Since $\beta \Delta p=\pi/{\sqrt{6n_{\rm{max}}}}$, the continuum limit is $n_{\rm{max}}\rightarrow \infty$, or $L\rightarrow \infty$ keeping temperature fixed. In this limit we can approximate the sum by the integral
\begin{eqnarray}\label{Large-L-Entropy}
S_A(n) &=&\frac{1}{\beta\Delta p}\int_{0}^{\beta p_n}{\rm d}x\,\left( x \frac{e^{-x}}{1-e^{-x}}-\log(1-e^{-x})\right) +{\cal O}(1/L)\\
&=&\frac{S}{2}-\frac{\pi n^2}{6S}-\frac{1}{\beta\Delta p}\left(
Li_2(1-e^{\beta p_n})+Li_2(e^{-\beta p_n})\right) +{\cal O}(1/L)\nonumber\ ,
\end{eqnarray}
where we used the indefinite integral
\begin{equation}
\int\,{\rm d}x\,\left( x \frac{e^{-x}}{1-e^{-x}}-\log(1-e^{-x})\right) = -\frac{x^2}{2}-Li_2(1-e^x)-Li_2(e^{-x})\ ,
\end{equation}
and $S$ is the total entropy \eqref{totalentropy}. As a crosscheck one can verify that in the limit $n\rightarrow \infty$, one obtains the total entropy. This can either be seen from the asymptotic expansion of the dilogarithm, or by directly doing the integral
\begin{equation}
S=\frac{L}{\pi\beta}\int_{0}^{\infty}{\rm d}x\,\left( x \frac{e^{-x}}{1-e^{-x}}-\log(1-e^{-x})\right)\ .
\end{equation}
We can also determine the thermal entropy of the complementary system, which we denote by $B$, namely the system of momentum modes that lie between a given $n$ and $n_{\rm{max}}$. This is given by
\begin{equation}\label{Wilsonthermalentropy2}
S_B (n) \equiv \sum_{k=n}^{n_{\rm{max}}}S_\beta(k)=\sum_{k=n}^{n_{\rm{max}}}\left(\beta p_k\,  \frac{e^{-\beta p_k}}{1-e^{-\beta p_k}}-\log(1-e^{-\beta p_k})\right) \ .
\end{equation}
Applying the previous continuum limit we can obtain also analytical formulas for the entropy of $B$. The plot of the thermal entropy of subsystem $B$ is given in Figure \eqref{fig5}.
The thermal entropies associated to $A$ and $B$ have different functional structures as we vary the subsystem sizes. Given that the entanglement entropy of $A$ should be equal to that of $B$ it might naively seem there is a contradiction here. But indeed there is no contradiction, since the thermal  entropy is a good approximation for the entanglement entropy of $A$ for values up to $n=21$ (half the entropy), and for the entanglement entropy of $B$ for larger values of $n$.

Therefore, what is meaningful here is the critical momentum mode $p_{n}$ for which the thermal entropy of the reduced subsystem is precisely half of the total entropy. This happens when the corrections to the first term in \eqref{Large-L-Entropy} cancel between each other. Remarkably, in the large $L$ limit, there is an exact solution of this, due to the identity of dilogarithms
\begin{equation}
Li_2(1-x)+Li_2(1-x^{-1})=-\frac{1}{2}(\log x)^2\ ,
\end{equation}
which we will use for $x=2$ and $\beta p_n=\log 2$ in \eqref{Large-L-Entropy}. For these choices, all other terms cancel except for the first one. Hence the critical
momentum dividing the QFT in two halfs maximally entangled with each other is given by
\begin{equation}\label{crit-momentum}
p^{\textrm{crit}}=T\log 2\:,
\end{equation}
giving a different interpretation of temperature in the QFT. From this perspective, the temperature $T$ provides the energy scale wich divides the QFT into two equal parts maximally entangled with each other, with a entanglement entropy equal to $S/2$.

In terms of the mode numbers, \eqref{crit-momentum} can be written as
\begin{equation}
n^{\textrm{crit}}=\frac{{\sqrt 6}\log 2}{\pi}\,\sqrt{n_{\rm{max}}}\approx 0.54 \sqrt{n_{\rm{max}}}\ .
\end{equation}
This can be checked by looking at Figure \ref{fig3}, for which we have $n^{\textrm{crit}}=0.54 {\sqrt{1500}}\approx 21$. For this value of $n$, we have $S_A(21)=47.33$, whereas $S/2=49.67$, so indeed around half the total entropy.

\begin{figure}[h]
\centering
{\includegraphics[height=42mm]{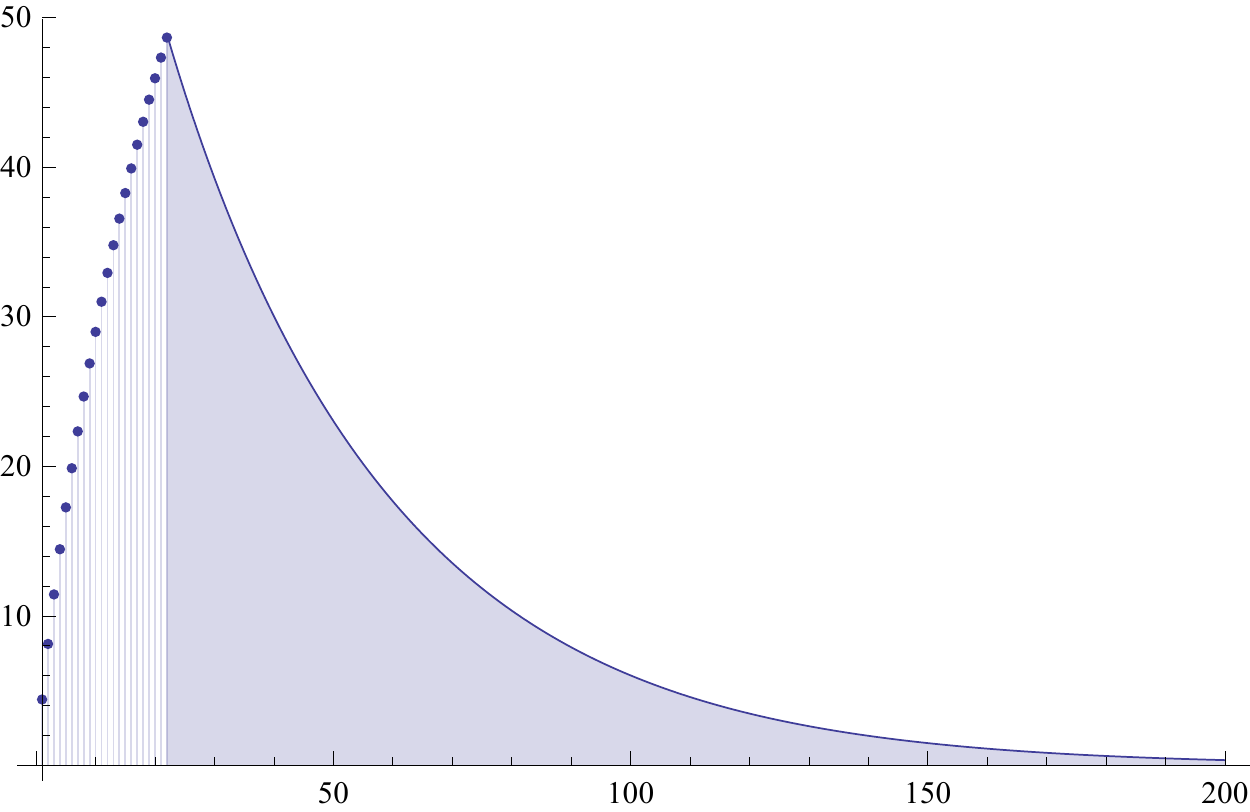}}
\caption{{\footnotesize The `Page curve', i.e the average entanglement entropy of reduced subsystems in random states, as a function of the energy scale dividing high and low energy degrees of freedom. The turning point, where the entanglement entropy takes its maximum value $S/2$, occurs for an energy scale of $\mathcal{O}(T)$, providing a different interpretation of temperature in the field theory.}}
\label{Page}
\end{figure}

The Page curve, as a function of the energy scale dividing high and low energy degrees of feedom, can now easily be found by combining the two curves in Figures \ref{fig3} and \ref{fig5}. They intersect exactly at $n^{\textrm{crit}}\approx 0.54 \sqrt{n_{\rm{max}}}$, as depicted in Figure \ref{Page}.

\subsection{The large N limit}

As commented in the introduction, part of the reasons to generalize the random unitary framework to the context of quantum field theory is to get closer to the physics of black holes\footnote{This formalism might be also interesting in the context of integrable theories and Generalized Gibbs Ensembles, see \cite{integrable}.}. In particular our formalism can be useful to describe black holes in anti de Sitter spacetimes via AdS/CFT \cite{maldacena}. In these theories we have a large gauge group, with a corresponding large number of field species. In AdS/CFT this number of field species is counted by the central charge $c$ of the CFT, which is taken to be large to have a smooth gravitational dual background. With this in mind we would like to repeat the previous exercise for the case of $N$ scalar fields, in the large $N$ limit. 

For each scalar field, labelled by $a=1,\cdots , N$, we have the same energy and momentum dispersion relation for excitations over the vacuum
\begin{equation}
E_{n_{a}}=p_{n_{a}}=\frac{\pi n_{a}}{L}\;,
\end{equation}
for a segment of length $L$ and $n_{a}=1,2,\cdots$.  The Fock space is spanned by states of the type
\begin{equation}\label{fockstatesN}
\vert \alpha\rangle\equiv\vert i_{n_{a}},i_{n_{b}},\cdots\rangle\;,
\end{equation}
where $i_{n_{a}}$ is the number of particles of the scalar field $a$ with momentum $p_{n_{a}}$, and $\alpha$ is again a natural number running over the infinite but countable set of eigenstates, used here to simplify notation. Conservation of total energy
\begin{equation}\label{energyscalarN}
E_{\textrm{T}}=\sum_{a=1}^{a=N}\sum_{n_{a}=1}^{n_{a}=n_{\textrm{max}}}i_{n_{a}} p_{n_{a}} = \sum_{a=1}^{a=N}\frac{\pi}{L}\sum_{n_{a}=1}^{n=n_{\textrm{max}}}i_{n_{a}} n_{a}\;,
\end{equation}
where $n_{\textrm{max}}$ is as in the one-field case $p_{n_{\textrm{max}}}=E_{\textrm{T}}$ (so $n_{\textrm{max}}=LE_{\textrm{T}}/\pi$), implies that any initial state with definite total energy $E_{\textrm{T}}$ will evolve towards states of the form
\begin{equation}
\vert\Psi_{\textrm{out}}\rangle=\sum_{\alpha=1}^{\alpha=\Omega (E_{\textrm{T}},N)}\Psi_{\alpha}\vert \alpha\rangle\;,
\end{equation}
where $\vert \alpha\rangle$ run over all the states belonging to (\ref{fockstatesN}) with total energy $E_{\textrm{T}}$. By definition these are
$\Omega (E_{\textrm{T}},N)=e^{S (E_{\textrm{T}},N)}$, where $S (E_{\textrm{T}},N)$ is the microcanonical entropy at energy $E_{\textrm{T}}$ of the theory with $N$ scalar fields. In Appendix~\ref{app2} this number is computed, and in the limit $E_{\textrm{T}}\gg N\pi/L$ or equivalently, $n_{\textrm{max}}\gg N$, we obtain
\begin{equation}
\Omega (E_{\textrm{T}},N)=\left(p(\frac{n_{\textrm{max}}}{N})\right)^{N}\rightarrow \left(\frac{N}{4n_{\textrm{max}}\sqrt{3}}\right)^{N}e^{2\pi\sqrt{\frac{n_{\textrm{max}}N}{6}}}\:,
\end{equation}
consistent with the Cardy formula for the microcanonical entropy:
\begin{equation}
S(E_{\textrm{T}},N)\simeq \sqrt{\frac{2\pi}{3} LE_{\rm{T}}N}=2\pi\sqrt{\frac{1}{6}c\, \Delta}\ ,\qquad
\Delta=L_0-\frac{c}{24}=n_{\textrm{max}}\qquad c=N\ .
\end{equation}
Now we can repeat the same generic procedure described in Section~\ref{genericprocedure}. In this case we will apply the generic formulas for the average density matrix and entanglement entropy~(\ref{typicalreducedsubset}) and (\ref{entanglementsubset}) directly. The average of the global density matrix is given by
\begin{equation}
[\rho_{\textrm{out}}]=[\vert \Psi_{\textrm{out}}\rangle  \langle \Psi_{\textrm{out}}\vert]=\sum_{\alpha=1}^{\alpha=\Omega (E_{\textrm{T}},N)}\frac{1}{\Omega (E_{\textrm{T}},N)}\vert\alpha\rangle\langle\alpha\vert\;,
\end{equation}
which is again the microcanonical density matrix.

The typical density matrix associated to a definite momentum $p_{n_{a}}$ is
\begin{equation}\label{density2N}
[\rho_{n_{a}}]=\sum_{i_{n_{a}}=0}^{i^{\textrm{max}}_{n_{a}}}\frac{\Omega (E_{\textrm{T}},N,i_{n_{a}})}{\Omega (E_{\textrm{T}},N)}\vert i_{n_{a}}\rangle\langle i_{n_{a}}\vert=\sum_{i_{n_{a}}=0}^{i^{\textrm{max}}_{n_{a}}}P(E_{\textrm{T}},N,i_{n_{a}})\vert i_{n_{a}}\rangle\langle i_{n_{a}}\vert\;,
\end{equation}
where $\Omega (E_{\textrm{T}},N,i_{n_{a}})$ is the number of states with total energy $E_{\textrm{T}}$ and $i_{n_{a}}$ particles with momentum $p_{n_{a}}$, and $i^{\textrm{max}}_{n_{a}}=\lfloor E_{\textrm{T}}/p_{n_{a}}\rfloor$. Therefore, the average entanglement entropy of a particle with momentum $p_{n_{a}}$ is given by
\begin{equation}\label{entropy2N}
[S_{\textrm{E}}(\rho_{n_{a}})]=-\sum_{i_{n_{a}}=0}^{i^{\textrm{max}}_{n_{a}}}P(E_{\textrm{T}},N,i_{n_{a}})\log P(E_{\textrm{T}},N,i_{n_{a}})\;.
\end{equation}
Formulas \eqref{density2N} and \eqref{entropy2N} are again the analogues of \eqref{typicaldensity} and \eqref{entropy} for the case at hand. To analyze the previous formulas we need to find the constrained multiplicities $\Omega (E_{\textrm{T}},N,i_{n_{a}})$. This is computed in Appendix~\ref{app2} in the same limit $E_{\textrm{T}}\gg N\pi/L$ as before. The result reads
\begin{equation}
\Omega (E_{\textrm{T}},N,i_{n_{a}})=\left(p(\frac{_{\textrm{max}}-i_{n_{a}}n_{a}}{N})\right)^{N}-\left(p(\frac{n_{\textrm{max}}-n_{a}(i_{n_{a}}+1)}{N})\right)^{N}\;,
\end{equation}
where we remind that $n_{\textrm{max}}=LE_{\textrm{T}}/\pi$. Expanding the probabilities $P(E_{\textrm{T}},N,i_{n_{a}})$ for $i_{n_{a}}n_{a}\ll n_{\textrm{max}}$, the leading term is given by
\begin{equation}
P(E_{\textrm{T}},N,i_{n_{a}})\rightarrow e^{-\beta i_{n_{a}} p_{n_{a}}}-
e^{-\beta p_{n_{a}}(i_{n_{a}}+1)}\;,
\end{equation}
which is just the Gibbs ensemble at temperature   
\begin{equation}
T=\frac{1}{\beta}=\sqrt{\frac{6 E_{\textrm{T}}}{\pi N L}}\ .
\end{equation}
Notice that again $TS=2E_{\textrm{T}}$, so that the pressure density reads
\begin{equation}
{\cal{P}}=E_{\textrm{T}}/V=\epsilon\;,
\end{equation}
 as expected for a two-dimensional conformal field theory.

We can now compute the deviations form thermality, to check if there is some extra dependence on the central charge $c=N$ of the theory. The next to leading term in the previous expansion is given by:
\begin{equation}
P(E_{\textrm{T}},N,i_{n_{a}})=P(T,N,i_{n_{a}})+N\frac{i_{n_{a}} p_{n_{a}}}{E_{\textrm{T}}}e^{-\beta i_{n_{a}} p_{n_{a}}}-N\frac{p_{n_{a}}(i_{n_{a}}+1)}{E_{\textrm{T}}}e^{-\beta p_{n_{a}}(i_{n_{a}}+1)}\;.
\end{equation}
The error we obtain is again much bigger than usually considered. For a typical momentum mode with energy of $\mathcal{O}(T)$, the $\textrm{error}^{N}_{E_{\textrm{T}}}(T)$ is:
\begin{equation}\label{errorN}
\textrm{error}^{N}_{E_{\textrm{T}}}(T)\sim\frac{N T}{E_{\textrm{T}}}\sim\frac{N}{S(E_{\textrm{T}},N)}\sim \frac{1}{TL}\;.
\end{equation}
In conformal field theory language, this translates into
\begin{equation}\label{errorCFT}
\textrm{error}^{c}_{\Delta}(T)\sim \sqrt{\frac{c }{\Delta}}\sim \frac{1}{TL}\ ,
\end{equation}
so for large conformal dimensions $\Delta$ compared to the central charge $c$, the errors are small. Equivalently, for high temperatures, the errors are small.

There is a simple intuition behind this result, given the approximation in which we are doing the computations. As explained in Appendix~\ref{app2}, to compute the microcanonical degeneracy $\Omega (E_{\textrm{T}},N)$ for the case of $N$ fields, we need to perform a difficult sum over different partitions. To compute it, we use a saddle point approximation. The physics behind the saddle point approximation, expected to be valid for $E_{\textrm{T}}\gg N\pi/L$, is that the problem at hand is equivalent to $N$ single scalar field theories, each one with a total energy $\bar{E}_{\textrm{T}}=E_{\textrm{T}}/N$. In this approximation we can then use the results of the single field model with total energy given by $\bar{E}_{\textrm{T}}$. Formula~(\ref{error}) then provides an error of order $\frac{i_{n_{a}} p_{n_{a}}}{\bar{E}_{\textrm{T}}}=N\frac{i_{n_{a}} p_{n_{a}}}{E_{\textrm{T}}}$, and indeed, in the limit $E_{\textrm{T}}\gg N\pi/L$, the error is small.

We want to remark here that one should not believe~(\ref{errorN}) at all energies, since the previous formulas for the multiplicities are only valid in the limit $E_{\textrm{T}}\gg N\pi/L$, or equivalently, for high temperatures $TL\gg 1$. Naively it might seem that thermal physics should not be valid for $E_{\textrm{T}}\lesssim N\pi/L$, but this is not the case. For smaller energies it is difficult to do the exact computations. But to see that thermal dynamics is still a good approximation, and compute the deviations from it, we can consider the specific case in which the total energy is given by $E_{\textrm{T}}=\pi/L$. The only Fock states with such an energy are the ones with one particle excited in the lowest momentum mode. There are $N$ of such states, one per field specie. Therefore, using the previous random machinery, the typical reduced density matrix of the lowest momentum mode of a single field is given by
\begin{equation}
\rho (p_{1})=\frac{N-1}{N}\vert 0_{p_{1}}\rangle \langle 0_{p_{1}}\vert+\frac{1}{N}\vert 1_{p_{1}}\rangle\langle 1_{p_{1}}\vert\:.
\end{equation}
This is just a mixed state, with probability equal to $\frac{N-1}{N}$ for specie 1 to be in the vacuum and probability equal to $\frac{1}{N}$ of having its lowest momentum mode occupied. Matching with the associated Gibbs ensemble is not as transparent as before. One possibility is to match the first probability exactly, from which we obtain $\beta = \frac{L}{\pi}\log N$. With this effective temperature the other probability is given by $\frac{1}{N}(1-1/N)$. We see that the deviation from thermality is of $\mathcal{O}(1/N)$ in this specific case, so we conclude that thermal dynamics is still a good approximation to the expected unitary microscopic result in the large $N$ limit. At any rate, it seems there might be a qualitative difference in the analysis when going from energies $E_{\textrm{T}}\gg N\pi/L$ to $E_{\textrm{T}} \lesssim N\pi/L$, or equivalently from high temperatures to below a critical temperature $T_{\rm crit}\sim 1/L$. Assuming that our results would also be valid for a CFT on a circle, and the field theory at large $N$ is dual to a gravitational bulk with AdS$_3$ radius proportional to $L$, then this would signal the Hawking-Page transition \cite{hawkingpage}, seen in the field theory in the size of deviations from thermality. This might have interesting implications in the context of AdS/CFT \cite{witten}. The regime we are working in corresponds to the black hole phase. Indeed, the black hole corresponds to a highly excited state in the CFT, so we require a large conformal dimension $\Delta$. Furthermore, large central charge is required for the bulk theory to be weakly coupled, and finally, we require $\Delta \gg c$ or $T\gg T_{\rm crit}$ to be well above the Hawking-Page transition point.

Moving forward, the average entanglement entropy of the single momentum mode $n_a$ is given by
\begin{equation}\label{ent-entN}
S_{\textrm{E}}(n_{a},n_{\textrm{max}},N)=-\sum_{i_{n_{a}}=0}^{i_{n_{a}}^{\textrm{max}}}P(E_{\textrm{T}},N,i_{n_{a}})\log P(E_{\textrm{T}},N,i_{n_{a}})\:.
\end{equation}
As stated before, this is a complicated sum for which we do not have a definite analytic expression. For the case of $n_{a}=n_{\textrm{max}}$ one can explicitly do the sum, since there are only two terms, $i_{n_{a}}=0$ and $i_{n_{a}}=1$. The result is
\begin{eqnarray}
S_{\textrm{E},N}(n_{\textrm{max}})&=&-\left(1-\frac{1}{p(n_{\textrm{max}},N)}\right) \log\left(1-\frac{1}{p(n_{\textrm{max}},N)}\right)\nonumber\\
&&-\frac{1}{p(n_{\textrm{max}},N)} \log\frac{1}{p(n_{\textrm{max}},N)}
\approx S(\textrm{E},N)\,e^{-S(\textrm{E},N)}\:,
\end{eqnarray}
where in the second line we took the leading term in the limit $n=n_{\textrm{max}}\rightarrow \infty$. We see that including more field species does not spoil the appearance of exponentially suppressed entanglement entropies in momentum space. We remark again that we find it very unprobable that geometric quantities, such as geodesic lengths,  can capture such physics, since the minimum proper lengths are of $\mathcal{O}(1)$ in Planck units.

Other aspects of the entropy behavior do not change drastically either from the single field case. This is easily observed by noticing that the thermal entropy of a given momentum mode, whatever the field specie, is left basically unchanged by the inclusion of more field species. The only change comes again by the total energy associated to a given field, and therefore the effective temeprature associated to each mode. For $N$ fields, the total energy given to one field is $\bar{E}_{\textrm{T}}=E_{\textrm{T}}/N$, and therefore the temperature $T=\sqrt{\frac{6\bar{E}_{\textrm{T}}}{\pi L}}=\sqrt{\frac{6E_{\textrm{T}}}{\pi N L}}$ decreases with $N$ as well. The deviations are given again by $\frac{i_{n_{a}} p_{n_{a}}}{\bar{E}_{\textrm{T}}}=N\frac{i_{n_{a}} p_{n_{a}}}{E_{\textrm{T}}}\sim N/S$, as the previous probabilities themselves.

Integrating out all momentum modes higher than a certain critical momentum $p_{n}$ is also possible. The associated constrained multiplicity can be computed in the same way as the previous ones, and an analogous formula to~(\ref{probn}) can be obtained. Therefore, one can approximate the typical reduced density matrix with the Gibbs ensemble in the same previous limit. We will not repeat the procedure here, since it is straightforward and those results are not affected by the inclusion of more fields.

Besides, the analogous of Page curve in this QFT setup is not going to change either. To observe this, notice that up to subleading corrections in the thermodynamic limit, we can use the thermal entropy for the low-energy modes. This is simply the sum of the thermal entropies of the individual modes of every field specie
\begin{equation}\label{Wilsonthermalentropy-N}
S=\sum_{a=1}^{a=N}\sum_{k_{a}=1}^{n}S_\beta(k_{a})=N \sum_{k=1}^{n}\left(\beta p_k\,  \frac{e^{-\beta p_k}}{1-e^{-\beta p_k}}-\log(1-e^{-\beta p_k})\right) \;,
\end{equation}
where we remind that $\beta=\sqrt{\frac{\pi N L}{6 E_{\textrm{T}}}}$.  Due to the simplicity of the previous relation, the critical momentum $p_{n}$ for which the thermal entropy of the reduced subsystem is precisely half of the total entropy does not change when considering $N$ fields. Hence, the critical
momentum dividing the QFT in two halfs maximally entangled with each other is again
\begin{equation}\label{crit-momentum-N}
p^{\textrm{crit}}=T\log 2\:,
\end{equation}
giving a different interpretation to temperature in the QFT.

\section{Conclusions}

The use of random unitary dynamics is having a strong impact in the physics of black holes and in the context of quantum thermalization, see the nice recent reviews \cite{vijay,harlow} and references therein. Probably the most famous example comes when considering finite dimensional Hilbert spaces, and when using the Haar ensemble of random unitary matrices. This example was first studied in \cite{Lloyd}, in which the typical properties of reduced density matrices were derived. Later, in \cite{page}, their results were used to compute the average entanglement entropy of a given subsystem.

In this article we have taken some steps towards generalizing the previous random unitary framework, so that it can be applied to any physical system and phase of it\footnote{Similar approaches have been considered in \cite{vijaytypical,nima,barbon,vijaymatrices,vijayent} in the context of black holes. In particular \cite{nima} also considers the application of Haar ensembles to microcanonical subspaces.}. We formulated the random unitary framework in Fock space. If unitary evolution constrains the state of the system to live in a specific subspace of the Hilbert space, defined by some set of charges, the typical properties of the subspace can be studied using the simple Gaussian relations~(\ref{constrainedevo}) and \eqref{ensemble}. This Gaussian ensemble is not precisely the Haar ensemble, but deviations between both of them are exponentially suppressed. Indeed, just with those Gaussian relations we were able to rederive the old results of \cite{Lloyd,page} in Appendix~\ref{app1}.

Formulating the problem in Fock space makes the study of reduced dynamics very natural. Generically one can choose any subset of particles, charcaterized by several quantum numbers, and integrate out all the others. The main results of the article are equations~(\ref{typicalreducedsubset}) and (\ref{entanglementsubset}). These are generic formulas for the typical reduced density matrix and the typical entanglement entropy associated to it. They are valid for any desired subsystem in any theory. They are written just in terms of `constrained multiplicities' $\Omega (E_{\textrm{T}},Q_{\textrm{T}}^{r},i_{a},i_{b},\cdots ,i_{c})$, i.e. the number of configurations with a given set of constraints. In  this way we connect modern approaches to quantum thermalization concerning quantum information theory, and in particular quantum entanglement, with more traditional microstate counting apporaches.

The standard deviations of both formulas are computed in Appendix~\ref{app1}. For the case of entanglement entropy we developed a new method to compute the deviations. The method directly connects the differences between Page's curve and the exact thermal answer to the well-known Wigner's semicircle law for the eigenvalue statistics of the ensemble of Gaussian Unitary Matrices (GUE)\footnote{See \cite{haake} for a nice introduction to the field of random matrices.}, see Appendix~\ref{app1}. This is very satisfactory, since it connects in a transparent way the more traditional approach to quantum chaos provided by random matrices and gaussian ensembles, see \cite{haake} for a nice introduction to the subject, to more modern approaches based on entanglement entropy.

Finally we applied the generic formulas for a massless scalar field theory in a finite segment, and for its generalization to the case of $N$ field species.  In this context we arrived at the following results:
\begin{itemize}
\item We showed the emergence of Gibbs dynamics with the correct effective temperature as a function of the total energy of the state. Any desired temperature can be produced, just by varying the total energy. The equation of state of the QFT is seen as the typical macroscopic configuration of the underlying microscopic quantum state.
\item The typical density matrix is not equal to the thermal density matrix, but to the microcanonical density matrix. Both are seen to be equal just in the strict thermodynamic limit. This is a huge difference with the old model described in \cite{Lloyd}, in which typicality is equal to thermality.
\item The deviations from thermality are of $\mathcal{O}(1/S)$ instead of $\mathcal{O}(e^{-S})$. These are much bigger than expected, and it seems just due to energy conservation. This result might imply that information from black hole evaporation can be extracted within perturbation theory.
\item We find exponentially suppressed entanglement entropies in the random state. The smallest ones are of $\mathcal{O}(S e^{-S})$. In the context of holography, we do not expect this to be captured by geometrical bulk quantities, such as geodesic lengths, since these can only measure entanglement entropies with a minimum size of $\mathcal{O}(1)$.
\item Drawing the analogue of Page's curve in the QFT scenario, as a function of the energy scale, provides a sort of entanglement running for the random state. The critical momentum dividing the QFT in two halfs maximally entangled with each other is given by $p_{\textrm{crit}}=T\log2$, giving a different interpretation of temperature in the quantum theory.
\end{itemize}

We want to finish with some outlook. Since the generic equations \eqref{typicalreducedsubset} and\eqref{entanglementsubset} apply to any theory with a Fock basis structure, such as a QFT or certain spin systems, there are two interesting avenues for future research in this regard. 

The first one is to apply the formulas to integrable systems, and see whether Generalized Gibbs Ensembles can be reproduced naturally, and what are the deviations from them. This might ultimately provide some more insight into generic differences between integrable and non-integrable quantum theories. Besides, although Generalized Gibbs Ensembles seem to fail in some situations, see \cite{integrable}, the random unitary ensembles we used in this article may represent better the physics of the integrable system.

The second one is to apply the framework to holographic field theories, beyond what is done in Section 3.2. The previous formulas seem a rigorous framework to study entanglement between infrared and ultraviolet domains at finite temperature. An exciting possibility is that it could turn into a workable route to extract physics from the near horizon region of quantum black holes. Aspects of these near horizon regions might be encoded in the structure of the Fock space entanglement of the field theory. At any rate we expect this framework to give more insights into the connections between entanglement and quantum gravity \cite{ryu,vijayhole}.

\section*{Acknowledgements}

It is a pleasure to thank Jose Barbon and Vijay Balasubramanian for interesting discussions.
This work was supported by the Netherlands Organisation for Scientific Research (NWO) under the VICI grant 680-47-603, and the Delta-Institute for Theoretical Physics (D-ITP) that is funded by the Dutch Ministry of Education, Culture and Science (OCW). This work was also supported by COST Action MP1210 The String Theory Universe.

\newpage

\appendix

\section{Average entanglement entropy vs entanglement entropy of the average}\label{app1}

One of the main objectives of this article has been to give a generic expression for the average entanglement entropy of a given reduced subsystem in a random state. Mathematically we needed to compute the following average
\begin{equation}
S_{\textbf{A}}^{\textrm{typical}}=[S_{\textbf{A}}]=\int d\rho S_{\textrm{E}}^{\textbf{A}}=-\int d\rho \textrm{Tr}(\rho_{\textbf{A}}\log \rho_{\textbf{A}})\;.
\end{equation}
We will now develop a novel method to derive the average. We will show that the method is able to reproduce Page's result as a specific case, but it is otherwise generalizable to our QFT framework as well.

Let us consider first the classical analogue of the previous problem. This will not be just a pedagogical exercise, since we will need the results of this classical case for the more complicated quantum one.

Consider a random probability distribution $p_{i}$, for $i=1,\cdots ,n$. With this we mean that each entry $p_{i}$ is an independent random variable with certain probability distribution itself, with the only constraint of global probability normalization $\sum_{i}p_{i}=1$. The average over the randomness in the probability distribution itself will be termed by $[\cdots]$. For example, the moments of $p_{i}$ are
\begin{equation}
[p_{i}^{m}]\;,
\end{equation}
for $m=1,2,\cdots$. The mean will be denoted simply by $[p_{i}]=\bar{p_{i}}$. The fact that they are independent random variables implies
\begin{equation}
[p_{i}p_{j}]=[p_{i}^{2}]\delta_{ij}\;.
\end{equation}
Given that the mean of a probability distribution cannot be zero, it is more interesting to work with the random variable $\delta p_{i}=p_{i}-\bar{p_{i}}$, the moments of which can be easily obtained from the moments of $p_{i}$.

In this context, problems naturally appear when we want functions of $p_{i}$. The average value of a given function $f(p_{i})$, with respect to a single realization of $p_{i}$, will be represented as $[f(p_{i})]_{p}$
\begin{equation}
[f(p_{i})]_{p}=\sum_{i=1}^{i=n}p_{i}f(p_{i})\;.
\end{equation}
But since the probability distribution itself is a random variable, we have that $[f(p_{i})]_{p}$ is also a random variable, and we are more interested in its moments rather than in the variable itself. In particular the mean is given by
\begin{equation}
[[f(p_{i})]_{p}]=[\sum_{i=1}^{i=n}p_{i}f(p_{i})]=[\sum_{i=1}^{i=n}(\delta p_{i}+\bar{p_{i}})f(\delta p_{i}+\bar{p_{i}})]\;.
\end{equation}
To compute the average of the function we can proceed iteratively, by Taylor expanding the function in terms of the deviations from the mean. For example, in the case $f(p_{i})=-\log p_{i}$ we are computing the average of Shannon's entropy:
\begin{equation}
[f(p_{i})]_{p}=S(p)=-\sum_{i=1}^{i=n}p_{i}\log p_{i}\;,
\end{equation}
which can be Taylor expanded as
\begin{eqnarray}
S(p)&=&-\sum_{i=1}^{n}(\delta p_{i}+\bar{p_{i}})\log(\delta p_{i}+\bar{p_{i}}) \nonumber\\
&=&-\sum_{i=1}^{n}(\delta p_{i}+\bar{p_{i}})\left(\log(\bar{p_{i}})+\frac{\delta p_{i}}{\bar{p_{i}}}-\frac{1}{2}(\frac{\delta p_{i}}{\bar{p_{i}}})^{2}+\cdots \right)\nonumber \\
&=&-\sum_{i=1}^{n}\left(\bar{p_{i}}\log \bar{p_{i}}+\delta p_{i}(1+\log \bar{p_{i}})+\frac{1}{2}(\delta p_{i})^{2}\frac{1}{\bar{p_{i}}})+\cdots \right)\;.
\end{eqnarray}
Since $[\delta p_{i}]=0$ by construction, up to higher order terms in the Taylor expansion, the average Shannon's entropy is given by
\begin{equation}\label{average}
[S(p)]\simeq -\sum_{i=1}^{n}\bar{p_{i}}\log \bar{p_{i}}-\frac{1}{2}\sum_{i=1}^{n}\frac{1}{\bar{p_{i}}}[(\delta p_{i})^{2}]=S(\bar{p})-\textrm{DEA}\;,
\end{equation}
where $\textrm{DEA}$ stands for Deviation from the Entropy of the Average, a quantity that needs to be computed to make sure the first term is the leading term in the themodynamic limit, and for other potential applications as well.

We want to remark here, as it is more transparent now, that $\textrm{DEA}$ is not \emph{a priori} related to deviations from some definition of thermality. We want to make the case that there is a conceptual, precise and computable, difference between $\textrm{DEA}$, the Standard Deviations from Typicality $\sigma^{2} (S(p))$, and Deviations from Thermality $\textrm{DT}$. Indeed one could also be interested in the average deviation of a typical realization from the mean $[S(p)]$\footnote{This quantity has not been studied before in the context of quantum thermalization. We comment on this below.}. This is the standard deviation from typicality 
\begin{equation}
\sigma^{2} (S(p))\equiv [S(p)^2]-[S(p)]^{2}\;.
\end{equation}
To compute it, we need to find the first term in the sum using again the previous expansion method. We obtain
\begin{equation}
S(p)^2=S(\bar{p})^{2}-2S(\bar{p})\left(\sum_{i}\delta p_{i}(1+\log \bar{p_{i}})\right)+(\sum_{i}\delta p_{i} \log\bar{p_{i}})^{2}-S(\bar{p})\sum_{i}\frac{(\delta p_{i})^{2}}{\bar{p_{i}}}+\cdots\;.
\end{equation}
The average of this expression is given by
\begin{equation}
[S(p)^2]\simeq S(\bar{p})^{2}+\sum_{i}[(\delta p_{i})^{2}]\left((\log \bar{p_{i}})^{2}-\frac{S(\bar{p})}{\bar{p_{i}}}\right)\;,
\end{equation}
and combining formulas we finally obtain:
\begin{equation}\label{daverage}
\sigma^{2} (S(p))=\sum_{i}[(\delta p_{i})^{2}](\log \bar{p_{i}})^{2}+\cdots\;,
\end{equation}
where the dots indicate higher order terms in $\delta p$.

We have written the average entropy~(\ref{average}) and the standard deviation from it~(\ref{daverage}) in terms of $[(\delta p_{i})^{2}]$. If we could obtain this quantity for the problem of interest we could obtain the average quantities.

Let us come back now to the quantum case. As with the classical case we will call $[\rho]=\bar{\rho}$. For all cases discussed in this article we have $(\bar{\rho})_{ij}=(\bar{\rho^{*}})_{ij}=(\bar{\rho})_{ii}\delta_{ij}$, i.e the typical density matrix is a real and diagonal matrix. We follow the same procedure and change the integration variable, from the state $\rho$ to the deviation from the mean:
\begin{equation}
\rho= \delta \rho + \bar{\rho}\;,
\end{equation}
so by construction we have
\begin{equation}
[(\delta \rho)_{ii'}]=0\;,
\end{equation}
where $ii'$ are the matrix indices, and also
\begin{equation}
[(\delta \rho^{*})_{ii'}(\delta \rho)_{jj'}]=[(\rho^{*})_{ii'}(\rho)_{jj'}]-(\bar{\rho})_{ii}(\bar{\rho})_{jj}\delta_{ii'}\delta_{jj'}\;.
\end{equation}
We conclude that the reduced density matrix will be some typical density matrix plus some random matrix, with a structure that might depend on the case studied.

Finally, we can translate the average over $\rho$ to the average over its random eigenvalues $p_{i}$:
\begin{equation}
[S_{\textbf{A}}]=-\int d\rho \textrm{Tr}(\rho_{\textbf{A}}\log \rho_{\textbf{A}})=[-\sum_{i}p_{i}\log p_{i}]\;,
\end{equation}
and therefore, at first order, over the random eigenvalues $ \delta p_{i}$ of the random matrix $\delta\rho$:
\begin{equation}
[S_{\textbf{A}}]=[-\sum_{i}(\bar{p_{i}}+\delta p_{i})\log(\bar{p_{i}}+\delta p_{i})]\;.
\end{equation}
In this way, if we know the statistics of the random matrix $\delta\rho$ and the statistics of its eigenvalues, we can compute the average value of the entanglement entropy and its standard deviation using formulas~(\ref{average}) and (\ref{daverage}). We will compute it for bipartite systems next.

\subsection{Random dynamics, GUE ensembles and Page's formula}

Following Ref \cite{page, Lloyd} we consider a bipartite system, with total dimension $AB$, where $A$ is the dimension of subsystem $\textbf{A}$, and $B$ of subsystem $\textbf{B}$. If $\vert i\rangle$, with $i=1,\cdots ,A$, is a basis of $\textbf{A}$ and $\vert \alpha\rangle$, with $\alpha=1,\cdots ,B$, is a basis of $\textbf{B}$, the state of the system can generically be represented as
\begin{equation}
\vert\psi\rangle=\sum_{i,\alpha}\psi_{i\alpha}\vert i,\alpha\rangle
\end{equation}
The application of a random unitary to the previous state can be defined as in Section 2, except that we let it act in the full Hilbert space instead of the subspace of fixed energy. It reads
\begin{equation}\label{pagestat}
[\psi_{i\alpha}]=0 ~~~~~~~~ [\psi_{j \beta}^{*}\psi_{i\alpha}]=\Lambda\,\delta_{ij}\delta_{\alpha\beta}\;.
\end{equation}
At first order, the random unitary just produces $AB$ independent gaussian random variables with squared deviation given by $\Lambda$. To fix this deviation we just fix the average normalization of the state
\begin{equation}
[\langle \psi\vert\vert\psi \rangle]=\sum_{i,\alpha} [\psi_{i \alpha}^{*}\psi_{i\alpha}]=AB\Lambda=1\;.
\end{equation}
We can now trace out subsystem $\textbf{B}$. The reduced state for subsystem $\textbf{A}$ is given by
\begin{equation}
\rho^{\textbf{A}}=\sum_{i,j}(\sum_{\alpha}\psi^{*}_{j\alpha}\psi_{i\alpha})\vert i\rangle\langle j\vert\;.
\end{equation}
We can now derive the statistics of the reduced state from the previous statistics~(\ref{pagestat}). The average is given by
\begin{equation}
[\rho^{\textbf{A}}_{ij}]=\sum_{\alpha}[\psi^{*}_{j\alpha}\psi_{i\alpha}]=\frac{1}{A}\delta_{ij}\;,
\end{equation}
so the average density matrix is exactly equal to the thermal density matrix at infinite temperature. The mean probabilities are given by ${\bar p}_i=1/A$ for each $i$. We remark here that this is not generic, as we have shown in the article. When charges are conserved the average density matrix is naturally the microcanonical density matrix, and therefore it is close to the thermal density matrix, but not equal. In such cases computing the deviations from the entropy of the average is not as important, because the leading behavior of the deviations from thermality are just given by using the entropy of the average.

The deviation from the mean is seen to be
\begin{equation}
[(\rho^{\textbf{A}}_{ii'})^{*}\rho^{\textbf{A}}_{jj'}]=\sum_{\alpha,\beta}[\psi^{*}_{i\alpha}\psi_{i'\alpha}\psi_{j\beta}\psi^{*}_{j'\beta}]=\frac{1}{A^{2}}\delta_{ii'}\delta_{jj'}+\frac{1}{A^{2}B}\delta_{ij}\delta_{i'j'}\;.
\end{equation}
Therefore, the statistical properties of $\delta\rho^{\textbf{A}}=\rho-\bar{\rho^{\textbf{A}}}$ are just given by $[(\delta\rho^{\textbf{A}})_{ij}]=0$, and more significantly
\begin{equation}
[(\delta\rho^{\textbf{A}})^{*}_{ij}(\delta\rho^{\textbf{A}})_{kl}]=\frac{1}{A^{2}B}\delta_{ik}\delta_{jl}\;.
\end{equation}
We conclude that the statistics of $\delta\rho^{\textbf{A}}$ are those of a random matrix belonging to the Gaussian unitary ensemble (GUE), with $\sigma^{2}(\delta\rho^{\textbf{A}})=\frac{1}{A^{2}B}$. Given this insight we can use one of the basic results in random matrix theory, see \cite{haake} for example, which is the single eigenvalue probability distribution $f(\lambda)$. The function $f(\lambda)$ is the celebrated Wigner's semicircle law. For a matrix of size $N$, with deviation $\sigma$ for each of the entries, it is given by
\begin{equation}
f(\lambda)=\frac{2}{4\pi N \sigma^{2}}\sqrt{4N\sigma^{2}-\lambda^{2}}\;,
\end{equation}
so that $[\lambda]=0$ and $[\lambda^{2}]=N\sigma^{2}$. In our case $N=A$, and this implies that the eigenvalues of $\delta\rho^{\textbf{A}}$, given by $\delta p_{i}$, satisfy
\begin{equation}
[(\delta p_{i})^{2}]=\frac{1}{AB}\;.
\end{equation}
Using now relation~(\ref{average}) we obtain:
 \begin{equation}\label{averagepage}
[S(p)]=-\sum_{i=1}^{A}\bar{p_{i}}\log \bar{p_{i}}-\frac{1}{2}\sum_{i=1}^{A}\frac{1}{\bar{p_{i}}}[(\delta p_{i})^{2}]=S(\bar{p})-\frac{A}{2B}\;,
\end{equation}
which is, to the level of accuracy in which we are working, Page's result for the average entanglement entropy \cite{page}. The deviations from the entropy of the average are then
\begin{equation}\label{DEApage}
\textrm{DEA}=\frac{A}{2B}\;.
\end{equation}
When $A\simeq {\mathcal{O}}(1)$, the DEA is inversely proportional to the size of $\textbf{B}$, or equivalently, exponentially suppressed in the entropy $S=\ln(AB)$. On the other hand, when $A\simeq B$, the DEA is of ${\mathcal{O}}(1)$.

In this case, because the mean density matrix is exactly equal to the thermal density matrix, these $\textrm{DEA}$ deviations are equal to the deviations from thermality $\textrm{DEA}=\textrm{DT}=\frac{A}{2B}$, a feature which will not extend to the QFT case.

A nice feature of this framework is that it can be immediately extended to compute the deviations from typicality, which are given by relation~(\ref{daverage})
\begin{equation}
\sigma^{2} (S(p))=\sum_{i}[(\delta p_{i})^{2}](\log \bar{p_{i}})^{2}+\cdots=\frac{(\log A)^{2}}{B}\;.
\end{equation}
The standard deviation from typicality is therefore parametrically smaller than the previous $\textrm{DEA}=\frac{A}{2B}$.

\subsection{Average entanglement in a QFT framework}

The extension of the previous framework to the QFT case does not entail any more conceptual insights. It just brings some technical complexity, because the structure of the reduced density matrix is not as simple as the previous case. But the final result turns out to have the same form, i.e we will obtain $\textrm{DEA}=\frac{A}{2B}$ for an appropriate $A$ and $B$ that we determine below. What is different is that for the QFT case we have $\textrm{DEA}\neq \textrm{DT}$, since the typical reduced density matrix is not equal to the thermal one. Therefore we rigorously justify the use of the entropy of the average through the article, since it provides the leading difference from thermality in scenarios in which several charges are conserved.

Let us first consider the case of a reduced density matrix for one particle type $a$. This case is particularly easy since the reduced density matrix is already diagonal before averaging, see \eqref{density}. The diagonal entries, or the probabilities, are given by
\begin{equation}
p_{i_a}=\sum_{\alpha}\psi_{i_a,\alpha}\psi^*_{i_a,\alpha}\ .
\end{equation}
We can therefore compute the DEA directly from \eqref{average}. One easily finds
\begin{equation}
{\bar p}_{i_a}=\frac{\Omega(E_{\rm T},i_a)}{\Omega(E_{\rm T})}\ ,\qquad [(\delta p_{i_a})^2]=\frac{\Omega(E_{\rm T},i_a)}{\Omega(E_{\rm T})^2}\ .
\end{equation}
The deviation from the entropy of the average is then
\begin{equation}
{\rm DEA}_a=\frac{1}{2}\frac{(i_a^{\rm max}+1)}{\Omega(E_{\rm T})}=\frac{1}{2}(i_a^{\rm max}+1)e^{-S}\ ,
\end{equation}
hence exponentially suppressed in the entropy $S$.

Now we generalize to any desired subset of particles $a,b,\cdots,c$. This is more difficult, since the reduced density matrix is no longer diagonal before averaging. To derive the deviations from the entropy of the average we have to understand the structure of the reduced density matrix. The reduced density matrix is given by formula~(\ref{reducedsubset}), which we rewrite here
\begin{eqnarray}\label{reducedsubset2}
\rho_{a,b,\cdots ,c}&=&\sum_{\alpha}\langle \alpha\vert  \Psi_{\textrm{out}}\rangle\langle\Psi_{\textrm{out}}\vert \alpha\rangle \\
&=& \sum_{i,i'} \left(\sum_{\alpha}\Psi_{i_{a},i_{b}, \cdots ,i_{c},\alpha}\Psi^{*}_{i'_{a}, i'_{b},\cdots , i'_{c}, \alpha}\right)\vert i_{a},i_{b},\cdots ,i_{c}\rangle\langle i'_{a},i'_{b},\cdots ,i'_{c}\vert\ ,\nonumber
\end{eqnarray}
where $\alpha$ labels the set of particles which are traced out. Although this is not a diagonal density matrix we notice that it has a certain block diagonal structure, due to energy and charge conservation. The entries outside the blocks are direcly zero, without the need of averaging. The blocks are defined by their total energy $E_{\textrm{T}}^{i_{a},i_{b},\cdots ,i_{c}}$ and their total charge $Q_{\textrm{T}}^{i_{a},i_{b},\cdots ,i_{c}}$. The dimension of each block is, by definition, $\Omega (E_{\textrm{T}}^{i_{a},i_{b},\cdots ,i_{c}}, Q_{\textrm{T}}^{i_{a},i_{b},\cdots ,i_{c}})$, another multiplicity that can in principle be computed\footnote{The generating function of such a multiplicity is indeed easy to write with the techniques described in Appendix~\ref{app2}. But we will not need its exact expression in a given theory, as we show below.}.

The statistics of the reduced density matrix are its average
\begin{equation}
\bar{\rho}_{a,b,\cdots ,c}\equiv[\rho_{a,b,\cdots ,c}]=\sum_{i}\frac{\Omega (E_{\textrm{T}},Q_{\textrm{T}}^{r},i_{a},i_{b},\cdots ,i_{c})}{\Omega (E_{\textrm{T}},Q_{\textrm{T}}^{r})}\vert i_{a},i_{b},\cdots ,i_{c}\rangle\langle i_{a},i_{b},\cdots ,i_{c}\vert\;,
\end{equation}
and the deviation from the mean, whis is given by:
\begin{equation}
[(\rho)_{ii'}^{*}(\rho)_{jj'}]=\frac{\Omega (E_{\textrm{T}},Q_{\textrm{T}}^{r},i)^{2}}{\Omega (E_{\textrm{T}},Q_{\textrm{T}}^{r})^{2}}\delta_{ii'}\delta_{jj'}+\frac{\Omega (E_{\textrm{T}},Q_{\textrm{T}}^{r},i)}{\Omega (E_{\textrm{T}},Q_{\textrm{T}}^{r})^{2}}\delta_{ij}\delta_{i'j'}\;,
\end{equation}
where we have used a short hand notation for $i\equiv i_{a},i_{b},\cdots ,i_{c}$ , etc, and where it is implicitly assumed that $i,i',j,j'$ belong to the same block, since in a different case the entries are directly zero.

From the previous expressions we can compute the statistics of $\delta\rho_{a,b,\cdots ,c}=\rho_{a,b,\cdots ,c}-\bar{\rho}_{a,b,\cdots ,c}$. By construction $[\delta\rho_{a,b,\cdots ,c}]=0$. The deviations are given by:
\begin{equation}
[(\delta\rho_{a,b,\cdots ,c})_{ii'}^{*}(\delta\rho_{a,b,\cdots ,c})_{jj'}]=\frac{\Omega (E_{\textrm{T}},Q_{\textrm{T}}^{r},i_{a},i_{b},\cdots ,i_{c})}{\Omega (E_{\textrm{T}},Q_{\textrm{T}}^{r})^{2}}\delta_{ij}\delta_{i'j'}\;.
\end{equation}
We conclude that $\delta\rho_{a,b,\cdots ,c}$ is a block diagonal matrix, with each block $B_{i}$ characterized by its total energy $E_{\textrm{T}}^{i_{a},i_{b},\cdots ,i_{c}}$ and its total charge $Q_{\textrm{T}}^{i_{a},i_{b},\cdots ,i_{c}}$, and where each block is a random matrix taken from GUE with size $S_{B_{i}}=\Omega (E_{\textrm{T}}^{i_{a},i_{b},\cdots ,i_{c}}, Q_{\textrm{T}}^{i_{a},i_{b},\cdots ,i_{c}})$ and deviation $\sigma^{2}_{B_{i}}=\frac{\Omega (E_{\textrm{T}},Q_{\textrm{T}}^{r},i_{a},i_{b},\cdots ,i_{c})}{\Omega (E_{\textrm{T}},Q_{\textrm{T}}^{r})^{2}}$.

If $\delta p_{i}$ are the eigenvalues of the block $B_{i}$, the contribution to the deviation from the entanglement entropy of the average of $B_{i}$ is (see formula~(\ref{average})) 
\begin{equation}
\textrm{DEA}_{B_{i}}=\frac{1}{2}\sum_{m\in B_{i}}\frac{1}{\bar{p}_{i}}[(\delta p_{i})^{2}]=\frac{1}{2}\frac{S_{B_{i}}^{2}}{\Omega (E_{\textrm{T}},Q_{\textrm{T}}^{r})}\;.
\end{equation}
The total deviation between the average entropy and the entropy of the average is finally
\begin{equation}\label{DEAQFT}
\textrm{DEA}=\frac{1}{2}\frac{\sum_{i}S_{B_{i}}^{2}}{ \Omega (E_{\textrm{T}},Q_{\textrm{T}}^{r})}\;.
\end{equation}
Although this expression seems opaque at first sight, it is exactly the same expression as for the Page's case. Notice that $\sum_{i}S_{B_{i}}^{2}$ counts the number of non-zero terms in the reduced density matrix, which for one particle subsystems is just $i_a^{\rm max}+1$. There is an analogous expression for the reduced density matrix of the complementary subsystem, say $\sum_{\bar{i}}S_{B_{\bar{i}}}^{2}$. The direct product of these two numbers must be the total non-zero entries of the full density matrix, so we have $\sum_{i}S_{B_{i}}^{2}\sum_{\bar{i}}S_{B_{\bar{i}}}^{2}=\Omega (E_{\textrm{T}},Q_{\textrm{T}}^{r})^{2}$. We conclude that $\sqrt{\sum_{i}S_{B_{i}}^{2}}=A$ provides the right dimension of the chosen subsystem, while for the complementary we have $\sqrt{\sum_{\bar{i}}S_{B_{\bar{i}}}^{2}}=B$. Relation~(\ref{DEAQFT}) is therefore exactly equivalent to relation ~(\ref{DEApage}). Finally, the deviations from typicality can be computed using~(\ref{daverage}) in a similar fashion.

We conclude that the difference between the average entropy and the entropy of the average density matrix behaves in the same way as the common Page's case \cite{page}, i.e it is of $\mathcal{O}(e^{-S})$ for small subsystems with dimensions of $\mathcal{O}(1)$, and it is of $\mathcal{O}(1)$ for subsystems with dimensions of $\mathcal{O}(S)$.

On the other hand, the deviations from thermality are naturally defined by the difference between the thermal entropy and the average entanglement entropy
\begin{equation}
\textrm{DT}\equiv S_{\beta}-[S_{E}]\simeq S_{\beta}-S_{\bar{\rho}}+\textrm{DEA}\ .
\end{equation}
In Page's case, the average reduced density matrix is equal to the thermal density matrix, and therefore the first two terms cancel each other, leaving $\textrm{DT}=\textrm{DEA}=A/2B$. In the QFT case, since for small subsystems we already have $S_{\beta}-S_{\bar{\rho}}\sim\mathcal (1/S)$, we do not need to consider the subtle $\textrm{DEA}$, since the leading corrections are already accounted by the entropy of the average density matrix.

\section{Partitions and constrained partitions}\label{app2}

One of the main messages developed in the article concerns the direct connection between typical reduced dynamics and multiplicity counting. In particular the computation of reduced density matrices and their associated entanglement entropies boils down to the computation of microcanonical and  constrained degeneracies. In the specific cases considered in this article this turns out to be possible, and we provide the detailed process in this appendix.

In the case of the real scalar field on a lign segment, we need to compute $\Omega (E_{\textrm{T}})$, the number of states at energy $E_{\textrm{T}}$, and $\Omega (E_{\textrm{T}},i_{n})$, the number of states at energy $E_{\textrm{T}}$ with $i$ particles with momentum $p_{n}$. $\Omega (E_{\textrm{T}})$ was seen to be the number of different partitions $p(n)$ of the natural number $\frac{L}{\pi}E_{\textrm{T}}$, which is given asymptotically by

\begin{equation}\label{asymptotics2}
\Omega (E_{\textrm{T}})=p(\frac{L}{\pi}E_{\textrm{T}})\rightarrow \frac{\pi}{4 \sqrt{3} L E_{\textrm{T}}}e^{\sqrt{\frac{2\pi}{3}E_{\textrm{T}}L}}\;.
\end{equation}
In the same way, $\Omega (E_{\textrm{T}},i_{n})$ is the number of different partitions $p(n',i_{n})$, of the natural number $n'=\frac{L}{2 \pi}E_{\textrm{T}}$, in which $n$ appears $i_{n}$ times. To compute it, notice the following theorem due to Euler:
\begin{equation}
P(x)=\frac{1}{1-x}\frac{1}{1-x^{2}}\frac{1}{1-x^{3}}\cdots\frac{1}{1-x^{i}}\cdots=\sum_{m=0}^{m=\infty}p(m)x^{m}\;,
\end{equation}
which provides a generating function for the number of different partitions $p(m)$. This is because
\begin{equation}
P(x)=(1+x+x^{2}+\cdots)(1+x^{2}+x^{4}+\cdots)(1+x^{3}+x^{6}+\cdots)\cdots (1+x^{n}+x^{2n}+\cdots)\cdots\;,
\end{equation}
and picking a monomial in the $k$-th part looks like $x^{i_k k}$ for arbitrary positive integers $i_k$. In the product $P(x)$, we get all monomials of the form $x^{i_11}x^{i_22}x^{i_33}\cdots=x^{i_1+2i_2+3i_3+\cdots}$ , which produce all possible partitions of the integer
\begin{equation}
m=i_1+2i_2+3i_3+\cdots n i_n\ ,
\end{equation}
with $n\leq m$ and $i_m=0$ or $1$.

To fix the number of times a given number $n$ appears in the partition, we just need to fix the monomial in the corresponding parenthesis, i.e we fix and select in the $n$-th bracket only the monomial $x^{i_nn}$. This way, we get the partition of $m$ in which the number $n$ appears $i_n$ times for fixed $(n,i_n)$. Hence, the generation function of $p(m,i_{n})$ is given by\footnote{The partition $p(m,i_n)$ should not be confused with the partition denoted by $p(m,n)$, which is the partition of $m$ with largest part $n$.}
\begin{eqnarray}\label{constraingenerating}
Q_{i_{n}}(x)&=&(1+x+x^{2}+\cdots)(1+x^{2}+x^{4}+\cdots)(1+x^{3}+x^{6}+\cdots)\cdots \nonumber\\
&\times& (1+x^{n-1}+x^{2(n-1)}+\cdots) x^{i_{n} n} (1+x^{n+1}+\cdots)\cdots\nonumber\\
&=&\sum_{m=0}^{m=\infty}p(m,i_{n})x^{m}\;.
\end{eqnarray}
In fact, the sum starts only from $m=i_n$ onwards, so $p(m,i_n)=0$ for $m<i_n$, which is also obvious from the definition of the partition.
The previous expression implies
\begin{equation}\label{constraingenerating2}
Q_{i_{n}}(x)=x^{i_{n} n}(1-x^{n})P(x)=\sum_{m=0}^{m=\infty}\left(p(m)x^{m+ i_{n} n}-p(m)x^{m+ n (i_{n}+1)}\right)\;,
\end{equation}
so we get, as long as $m-i_nn\geq n$,
\begin{equation}
p(m,i_n)=p(m-i_nn)-p(m-i_nn-n)\ .
\end{equation}
For $m-i_nn<n$, the identity is still true but without the second term. We now finally obtain
\begin{equation}
\Omega (E_{\textrm{T}},i_{n})=p(\frac{L}{ \pi}E_{\textrm{T}},i_{n})=p\left(\frac{L}{ \pi}E_{\textrm{T}}-i_{n} n\right)-p\left(\frac{L}{ \pi}E_{\textrm{T}}-n (i_{n} +1)\right)\;,
\end{equation}
a result that was used in Section 3.1. 

One of our main result in the text is the sub-leading corrections to the thermal entropy, given in \eqref{error}. It is easy to proof this formula using the Hardy-Ramanujan asymptotic formula for the number of partitions,
\begin{equation}\label{HR}
p(N)\approx \frac{1}{4\sqrt 3}\frac{1}{N}e^{\pi\sqrt{\frac{2N}{3}}}\ .
\end{equation}
Consider now two numbers $a$ and $b$, with $a<b$ and both $a\ll N$ and $b\ll N$. Define the combination
\begin{equation}
P(N,a,b)\equiv \frac{p(N-a)-p(N-b)}{p(N)}\ ,
\end{equation}
which is of the form of the reduced probabilities in \eqref{prob-one-cell}. Using \eqref{HR}, it is now an easy exercise to show that, in the $a/N$ expansion, we get
\begin{equation}\label{prob-expand}
\frac{p(N-a)}{p(N)}\approx (1+\frac{a}{N}+\cdots)\,e^{-\frac{\pi\,a}{\sqrt{6N}}\left(1+\frac{a}{4N}+\cdots\right)}\ .
\end{equation}
Higher order corrections are of order $a^2/N^2$ or more. One would be inclined to also drop the subleading term in the exponent, such that the only next to leading order term is the $a/N$ term in front of the exponent. This would be fine, unless $a\sim \sqrt N$, which is what we actually consider in the main text (when the momentum mode has energy of order $T$). In that case, $a/N\sim 1/{\sqrt N}$ is still small, but expanding the exponential now gives subleading terms that are of the same order as $a/N$. The terms don't cancel out each other but merely change the coefficient in front of $a/N$. So we conclude that
\begin{equation}\label{prob-error}
P(N,a,b)\approx \left(1+\mathcal{O}(\frac{a}{N})\right)\,e^{-\frac{\pi\,a}{\sqrt{6N}}}-\left(1+\mathcal{O}(\frac{b}{N})\right)\,e^{-\frac{\pi\,b}{\sqrt{6N}}}\ .
\end{equation}
The corrections are hence of order $a/N$, and if we take $a\sim \sqrt N$ (and similarly for $b$), the exponent is of order unity, and the total error scales like $1/\sqrt N$. In the main text, $1/\sqrt N \sim 1/S$, so the error is inversely proportional to the entropy.

Generalizing this procedure to include more constraints is straightforward. We could be interested in $\Omega (E_{\textrm{T}},i_{n},i_{l},\cdots)$, the number of states with $i_{n}$ units of momentum $n$, $i_{l}$ units of momentum $l$, etc. This is then equal to $p(m, i_{n},i_{l},\cdots)$ the number of partitions  in which $n$ appears $i_{n}$ times, $l$ appears $i_{l}$ times, etc, and where $m=\frac{L}{2 \pi}E_{\textrm{T}}$. We get this number from its corresponding generating function:
\begin{equation}
Q_{i_{n},i_{l},\cdots}(x)=x^{i_{n} n}(1-x^{n})x^{i_{l} l}(1-x^{l})\cdots P(x)\;.
\end{equation}

The previous formula can be applied for example to compute the entanglement between high energy momentum modes and low energy momentum modes. If we trace out momenta bigger than $p_{n}$, we need to analyze the following generating function:
\begin{equation}
Q_{i_{1},i_{2},\cdots ,i_{n}}(x)=x^{i_{1}}(1-x)x^{i_{2} 2}(1-x^{2})\cdots x^{i_{n} n}(1-x^{n})P(x)\;.
\end{equation}
To convert this to a generating function, we use the following identity:
\begin{equation}
(1-x)(1-x^{2})\cdots (1-x^{n})=1 -\sum_{l=1}^n x^l + \sum_{l<r}x^{l+r}-\sum_{l<r<s}x^{l+r+s}+\cdots+x^{1+2+\cdots+n}\ ,
\end{equation}
where e.g.
\begin{equation}
\sum_{l<r}=\sum_{l=1}^{n-1}\sum_{r=l+1}^n \ ,\qquad \sum_{l<r<s}=\sum_{l=1}^{n-2}\sum_{r=l+1}^{n-1}\sum_{s=r+1}^n\ .
\end{equation}
This provides a very complicated expression for the number of modes with $i_{1}$ particles with momentum $p_{1}$,$i_{2}$ particles with momentum $p_{2}$, etc., which is given by:
\begin{eqnarray}\label{lownumber}
\Omega (E_{\textrm{T}},i_{1},i_{2},\cdots ,i_{n})&=& p\left(\frac{L}{2 \pi}E_{\textrm{T}}-\sum_{k=1}^{n}i_{k} k\right)+\nonumber\\
&-&\sum_{l=1}^{n}p\left(\frac{L}{2 \pi}E_{\textrm{T}}-\sum_{k=1}^n i_{k} k-l\right)+ \nonumber \\
&+&\sum_{l<r}p\left(\frac{L}{2 \pi}E_{\textrm{T}}-\sum_{k=1}^n i_{k} k-l-r\right)+\cdots \nonumber\\
&+&(-1)^{n}p\left(\frac{L}{2 \pi}E_{\textrm{T}}-\sum_{k=1}^n i_{k} k-\sum_{l=1}^n l\right) \ .
\end{eqnarray}
Although this seems an opaque expression, it can be matched exactly with predictions from the Gibbs distribution, as was explained in a previous section.

Now we describe the computations in the second case, a theory of $N$ scalar fields. We will proceed in much the same fashion as the previous case. First we compute the generating function $Q^{N}(x)$ of the number of ways $p^{N}(n)$, with $n=LE_{\textrm{T}}/\pi$, of writing:
\begin{equation}\label{energyscalar2}
n= \sum_{a=1}^{a=N}\sum_{n_{a}=1}^{n=n_{\textrm{max}}}i_{n_{a}} n_{a}\;,
\end{equation}
It is simple to observe that:
\begin{equation}
Q^{N}(x)=\sum_{n}p^{N}(n)x^{n}=(\frac{1}{1-x})^{N}(\frac{1}{1-x^{2}})^{N}(\frac{1}{1-x^{3}})^{N}\cdots(\frac{1}{1-x^{i}})^{N}\cdots=(P(x))^{N}
\end{equation}
This is just seen by expanding each of the fractions, and verifying that the products provide all terms in~(\ref{energyscalar}). Therefore we obtain:
\begin{equation}
p^{N}(n)=\sum_{m,r,\cdots ,l}p(m)p(r)\cdots p(l)\delta_{n,m+r+\cdots +l}\:,
\end{equation}
where the number of indices $m,r,\cdots ,l$ is equal to $N$, and $p(n)$ is the usual number of partitions. The previous sum can be evaluated by a saddle point approximation, when $n\gg N$. Physically, the saddle point approximation is just the statement that with overwhelming probability the total energy $E_{\textrm{T}}\gg N\pi/L$ will be equally distributed over all $N$ scalar fields. Mathematically we have $m=r=\cdots =l=n/N$, and the previous degeneracy reads:
\begin{equation}
p^{N}(n)\simeq (p(n/N))^{N}\rightarrow \lambda^{N}e^{2\pi\sqrt{\frac{nN}{6}}}\:,
\end{equation}
where $\lambda=\frac{N}{4n\sqrt{3}}$. In this $n\gg N$ limit, the entropy now provides the usual Cardy formula:
\begin{equation}
S\simeq 2\pi\sqrt{\frac{1}{6}c n_{\textrm{max}}}\:,
\end{equation}
Finally, the generating function $Q_{i_{q_{a}}}^{N}(x)$ of the number of states of total energy $E_{\textrm{T}}$ and $i_{q_{a}}$ units of momentum $q_{a}$ associated to the field $a$ is given by 
\begin{equation}
Q_{i_{q_{a}}}^{N}(x)=\sum_{n}p^{N}_{i_{q_{a}}}(n)x^{n}=(\frac{1}{1-x})^{N}(\frac{1}{1-x^{2}})^{N}\cdots(\frac{1}{1-x^{q}})^{N-1}x^{i_{q_{a}}}(\frac{1}{1-x^{q+1}})^{N}\cdots\:,
\end{equation}
so that
\begin{eqnarray}
p^{N}_{i_{q_{a}}}(n)&=&\sum_{m,r,\cdots ,l}p(m)p(r)\cdots p(l)\delta_{n,m+r+\cdots +l+i_{q_{a}}q_{a}}- \nonumber \\ 
&-&\sum_{m,r,\cdots ,l}p(m)p(r)\cdots p(l)\delta_{n,m+r+\cdots +l+q_{a}(i_{q_{a}}+1)}\:.
\end{eqnarray}
In the previous $n\gg N$ limit, the saddle point approximation results in:
\begin{equation}
p^{N}_{i_{q_{a}}}(n)=\left(p(\frac{n-i_{q_{a}}q_{a}}{N})\right)^{N}-\left(p(\frac{n-q_{a}(i_{q_{a}}+1)}{N})\right)^{N}\:.
\end{equation}

%{toc}{chapter}
%\addcontentsline{toc}{section}{Bibliography}

%bibliography{bibmutual} % texto.bib es el fichero donde está salvada la bibliografía.
%\bibliographystyle{unsrt} % estilo de la bibliografía.

\end{document}